\documentclass[12pt]{article}
\usepackage{latexsym}
\usepackage{epsfig,amssymb,euscript,slashed}
\usepackage[linktocpage]{hyperref}
\usepackage{amsmath}
\usepackage[noadjust]{cite}
\usepackage{mathtools}
\usepackage{physics}
\usepackage{extarrows}
\usepackage{xcolor}
\usepackage{float}
\usepackage{graphicx}
\usepackage{subcaption}
\usepackage{tcolorbox}
\usepackage{relsize}
\usepackage{environ}
\usepackage{tikz}
\usepackage{array,calc,epsfig}
\usepackage{bbm}
\usepackage{fancybox}

\newcommand{\cM}{{\cal M}}

\numberwithin{equation}{section} 
\usepackage[]{graphicx}
\usepackage{color}

\allowdisplaybreaks
\begin{document}
\font\cmss=cmss10 \font\cmsss=cmss10 at 7pt

\begin{flushright}{  
\scriptsize QMUL-PH-21-49}
\end{flushright}
\hfill
\vspace{18pt}
\begin{center}
{\Large 
\textbf{AdS$_3$ holography for non-BPS geometries
}}

\end{center}

\vspace{8pt}
\begin{center}
{\textsl{Bogdan Ganchev$^{\,a}$, Stefano Giusto$^{\,b, c}$, Anthony Houppe$^{\,a}$ and Rodolfo Russo$^{\,d}$}}

\vspace{1cm}

\textit{\small ${}^a$  Université Paris-Saclay, CNRS, CEA, Institut de physique théorique,\\
91191, Gif-sur-Yvette, France.} \\  \vspace{6pt}

\textit{\small ${}^b$ Dipartimento di Fisica,  Universit\`a di Genova, Via Dodecaneso 33, 16146, Genoa, Italy.} \\  \vspace{6pt}

\textit{\small ${}^c$ I.N.F.N. Sezione di Genova,
Via Dodecaneso 33, 16146, Genoa, Italy.}\\
\vspace{6pt}

\textit{\small ${}^d$ Centre for Theoretical Physics, Department of Physics and Astronomy,\\
Queen Mary University of London,
Mile End Road, London, E1 4NS,
United Kingdom.}\\
\vspace{6pt}

\end{center}

\vspace{12pt}

\begin{center}
\textbf{Abstract}
\end{center}

\vspace{4pt} {\small
\noindent 
By using the approach introduced in~\cite{Ganchev:2021pgs} we construct non-BPS solutions of 6D $(1,0)$ supergravity coupled to two tensor multiplets as a perturbation of AdS$_3\times S^3$. These solutions are both regular and asymptotically AdS$_3\times S^3$, so according to the standard holographic framework they must have a dual CFT interpretation as non-supersymmetric heavy operators of the D1-D5 CFT. We provide quantitative evidence that such heavy CFT operators are bound states of a large number of light BPS operators that are mutually non-BPS. }

\vspace{1cm}

\thispagestyle{empty}

\vfill
\vskip 5.mm
\hrule width 5.cm
\vskip 2.mm
{
\noindent {\scriptsize e-mails: bogdan.ganchev@ipht.fr, stefano.giusto@pd.infn.it, anthony.houppe@ipht.fr,   r.russo@qmul.ac.uk }
}

\setcounter{footnote}{0}
\setcounter{page}{0}

\newpage

\tableofcontents


\section{Introduction}
\label{sec:intro}

In holographic Conformal Field Theories (CFTs) states with a conformal dimension of the order of the central charge may have a dual gravitational description as a smooth non-trivial geometry which is asymptotically AdS. There are several examples in supersymmetric theories where such heavy states are also BPS: for instance the half-BPS sector is well understood both in the D1-D5 CFT~\cite{Lunin:2002iz,Kanitscheider:2007wq} and in ${\cal N}=4$ Super-Yang Mills (SYM)~\cite{Lin:2004nb}. The underlying idea in both cases is that half-BPS smooth supergravity solutions are dual to bound states at threshold of many elementary constituents that preserve the same supercharges and correspond to Chiral Primary Operators (CPOs) on the CFT side.

In this paper we focus on the AdS$_3 \times S^3 \times {\cal M}_4$ setup, with $ {\cal M}_4$ being either $T^4$ or $K_3$ which is dual to the D1-D5 CFT. In the supergravity approximation the central charge of the CFT ($c=6N$) is large and smooth classical solutions describe coherent-like states where each species of CPOs appears many times, {\rm i.e.} a finite fraction of $N$. In this AdS$_3$/CFT$_2$ context substantial progress has also been made in the quarter BPS sector where~\cite{Bena:2015bea} initiated a systematic study of the geometries dual to heavy states made of BPS single-particle constituents that have in common only a quarter of the CFT supercharges. Again the idea is to focus on semiclassical states where the number of each species of constituents is large so that its ratio with respect to $N$ corresponds to a continuous parameter characterising the dual solutions. This class of BPS geometries is known as ``superstrata'', see~\cite{Bena:2016ypk,Bena:2017xbt,Ceplak:2018pws,Heidmann:2019zws,Heidmann:2019xrd} and~\cite{Shigemori:2020yuo} for a recent review. Superstrata are not the only supersymmetric asymptotically AdS solutions known in this context: another class of such solutions is provided by the so-called bubbling geometries~\cite{Warner:2019jll}, although their holographic interpretation is still not well understood.

It is natural to ask whether the picture sketched above can be generalised to describe semiclassical non-BPS states. A host of regular and horizonless non-supersymmetric geometries have been found by either generalising the bubbling BPS ansatz \cite{Bena:2015drs,Bena:2016dbw,Bossard:2017vii,Bah:2021jno} or by exploiting the Weyl formalism \cite{Bah:2021owp,Bah:2021rki} but, much like the bubbling geometries, these solutions do not currently have a known holographic dual. A first approach to the construction of non-supersymmetric microstates with a clear holographic interpretation is to consider non-BPS descendants of well understood BPS configurations as was done in~\cite{Jejjala:2005yu} or, at the perturbative level, in \cite{Bombini:2017got}. In the decoupling limit, this amounts to applying a large coordinate transformation and re-interpreting the resulting solution, while the extension to an asymptotically flat configuration requires an explicit analysis of the supergravity equation of motions, as was done for instance in \cite{Bombini:2017got}. In this work we will restrict ourselves to the asymptotically AdS setup, but we will study configurations that are not related by symmetries to known solutions. To remain within the realm of supergravity we consider heavy states whose individual constituents\footnote{In the D1-D5 CFT genuine non-BPS individual constituents are dual to string excitations that become infinitely heavy at the supergravity point.} are still BPS or descendants of BPS states, but that globally do not preserve any common set of supercharges: these states are the natural generalisation of the superstrata. We will work in perturbation theory by taking the number $N_i$ of each species of constituents to be a finite, but a small fraction of the central charge, so that $\alpha_i^2 \sim N_i/N\ll 1$ even if the $N_i$'s are large. Since we work in the large $N$ regime, the interaction between mutually non-BPS single-particle constituents is weak (it vanishes in the strict $N\to\infty$ limit) and then we can study quantitatively how the solutions behave when the parameters $\alpha_i$ are switched on. At linear order in $\alpha_i$ the non-BPS states we consider can be generated by applying a Virasoro generator to a supersymmetric state and the interactions are not present, but already at the quadratic order in $\alpha_i$  they display new features with respect to their supersymmetric relatives because the effects of the non-trivial interaction between pairs of constituents become visible. The main result of our analysis is that such semiclassical non-supersymmetric configurations are still described, to arbitrary order in perturbation theory\footnote{In the explicit examples we considered we checked this up to ${\cal O}(\alpha^{11})$.}, by {\em regular} supergravity solutions that are asymptotically AdS.

As expected, tackling interacting bound states is substantially more complicated than studying BPS configurations at threshold. Of course, on the bulk side, one needs to deal directly with the equations of motion rather than the simpler first order supersymmetry equations. A crucial step in addressing this challenge in our context has been provided by a recent series of works \cite{Mayerson:2020tcl,Houppe:2020oqp,Ganchev:2021pgs},which developed a new technique to construct regular asymptotically AdS$_3\times S^3$ supergravity solutions potentially dual to black hole microstates --  henceforth dubbed microstate geometries -- that does not rely on supersymmetry. The technique is based on a consistent $S^3$ reduction of 6D $(1,0)$ supergravity coupled to two tensor multiplets that was found in \cite{Samtleben:2019zrh}. The truncation of course simplifies the task of solving the equations of motion and makes the problem tractable even in the absence of supersymmetry: with some further assumptions, whose scope will be discussed in the following, the authors of \cite{Ganchev:2021pgs} have shown that the problem can be reduced to a system of {\it ordinary} differential equations, which can be solved perturbatively or numerically. The observation that one can look for microstate geometries inside a truncation is particularly non-trivial because a decennial body of work (see again \cite{Warner:2019jll,Shigemori:2020yuo} and references therein) has taught us that the existence of regular and normalisable\footnote{We use the term normalisable in the sense that has become familiar in the AdS/CFT context: a solution is normalisable if its fields decay at the asymptotic AdS boundary like a normalisable perturbation.} solutions requires keeping on the gravity side more degrees of freedom than the ones naively suggested by the classical black hole. Inputs from some microscopic description -- a worldsheet picture \cite{Giusto:2011fy} or a holographic CFT dual description \cite{Kanitscheider:2006zf,Giusto:2015dfa,Giusto:2019qig,Rawash:2021pik} -- have often been indispensable guides for the gravitational construction of the microstates. Since there is no guarantee that what is natural from a gravity perspective is also what is required by the microscopic analysis, the fact that the truncation of \cite{Samtleben:2019zrh} allows to describe a class of microstate geometries appears as a lucky circumstance. The first evidence in favour of this possibility was the realisation \cite{Mayerson:2020tcl} that a subset of the superstrata 
sits in the truncation. Then in~\cite{Ganchev:2021pgs} a class of non-supersymmetric solutions was studied both analytically -- in the perturbative approach mentioned above -- and numerically. Here we will build on this analysis and show how to construct regular, asymptotically AdS solutions dual to non-BPS states of the D1-D5 CFT. This is possible because the truncation contains some fields that are not excited in the superstrata but that should be used in order to construct the regular geometry dual to the CFT multi-particle heavy states that are not supersymmetric.

The paper is structured as follows. We start with a short review of the 3D truncation in Section~\ref{sec:trunCFT}. Emphasis will be given to the dual CFT interpretation deriving the explicit form of the single-particle modes that are encoded in the truncation. In Section~\ref{sec:sugra} we make use of this 3D approach to develop the supergravity analysis: we first review how the superstrata in~\cite{Bena:2016ypk} fit in the truncation and then derive two families of non-BPS solutions in the perturbative approach discussed above. In section~\ref{sec:holo} we focus on the physical properties of these solutions reading from their asymptotic behaviour the various conserved charges and discussing their holographic interpretation. In particular we show that the supergravity solutions are consistent with the picture sketched here as geometries describing multi-particle states composed by weakly interacting constituents. Section~\ref{sec:conclusions} contains a brief summary and a discussion of the challenges raised by the developments discussed in this paper. Three appendices provide some technical details on the ten-dimensional uplift of the six-dimensional solutions (Appendix~\ref{app:uplift10}), the linearised 3D equations of motion (Appendix~\ref{app:linearized_oem}), and the holographic map between supergravity fields and CFT operators (Appendix~\eqref{app:3d-CFT}). Supplemental material is provided in a zip, containing Mathematica files to reproduce the perturbative expansions and instructions how to use them.
 
\section{A 3D supergravity and its CFT interpretation}
\label{sec:trunCFT}

In this section we will clarify the holographic CFT interpretation of the field content of the truncation used in \cite{Mayerson:2020tcl,Houppe:2020oqp,Ganchev:2021pgs}, which we first briefly summarise. 

In its minimal form, which will be sufficient to describe the class of microstates we consider in this article, the ingredients defining the truncation are four scalars $\nu$, $\mu_0$, $\mu_1$, $\mu_2$, two 3D vectors $A^{\varphi_1}$, $A^{\varphi_2}$ and the 3D metric $ds^2_3$, which is in turn parametrised by the scalars $\Omega_0$, $\Omega_1$ and $k$. These quantities define a solution of 6D supergravity, which we identify with the reduction of type IIB supergravity on the 4D compact space $\mathcal{M}=T^4$ or K3. We parametrise the asymptotically AdS$_3$ space by the coordinates $\xi$, $\psi$ and $\tau$. The coordinate $\xi$ is related with the more conventional AdS$_3$ radial direction $\rho\in [0,\infty)$ via 
\begin{equation}
\xi=\frac{\rho}{\sqrt{1+\rho^2}}\,,
\end{equation}
so that $\xi\to 1$ corresponds to the AdS boundary; the coordinates $\tau$ and $\psi$ are related with the time and spatial coordinates, $t$, $y$, of the dual 2D CFT by 
\begin{equation}
\psi = \tau+\sigma\;,~~~~\tau=\frac{t}{R_y}\quad,\quad \sigma=\frac{y}{R_y}\,,
\end{equation}
where $R_y$ is the radius of the CFT spatial circle. The $S^3$ angles are $\theta\in[0,\frac{\pi}{2} ]$, $\varphi_1, \varphi_2\in [0,2\pi]$. In these coordinates, the  AdS$_3$ and the $S^3$ metrics with unit radius read
\begin{equation}
  \label{eq:ads3s3m}
  \begin{gathered}
    ds^2_{AdS_3}= \frac{d\xi^2}{(1-\xi^2)^2} -\frac{d\tau^2}{1-\xi^2}+\frac{\xi^2}{1-\xi^2}d\sigma^2 = -(\rho^2+1) d\tau^2 + \rho^2 d\sigma^2 + \frac{d\rho^2}{\rho^2+1}\,,
    \\
    ds^2_{S^3} = d\theta^2+\sin^2\theta d\varphi_1^2 + \cos^2\theta d\varphi_2^2\,,
      \end{gathered}
\end{equation}
and the orientation of these spaces is chosen so that their volume forms are
\begin{equation}
  \text{vol}_{AdS_3} ~=~ \frac{\xi}{(1-\xi^2)^2} d\xi \wedge dt \wedge d\psi \,,\qquad \text{vol}_{S^3} ~=~ \cos\theta \sin\theta \,d\theta \wedge d\varphi_1 \wedge d \varphi_2 \,.
\end{equation}

We also denote by $R_{AdS}$ the AdS$_3$ radius, which is linked to the D1 and D5 charges, $Q_1$, $Q_5$, of the IIB description by $R_{AdS}=(Q_1 Q_5)^{\frac{1}{4}}$. It will be enough for us to specify\footnote{The 6D solution contains also three self-dual three-forms $G^I$, which in the 10D description represent the RR and NSNS three-form field strengths, whose expressions can found in \cite{Mayerson:2020tcl}.} the 6D Einstein metric $ds^2_6$, the dilaton $\Phi$ and the RR zero-form $C_0$:
\begin{equation}\label{eq:metric6D3D}
\begin{aligned}
\frac{ds^2_6}{R_{AdS}^2} &= e^{-2(\mu_1+\mu_2)}\Delta^{\frac{1}{2}} ds^2_3\\
&+ \Delta^{-\frac{1}{2}}\left[e^{2\mu_1}\sin^2\theta +e^{2\mu_2}\left(e^{-2\mu_0} \cos^2(\varphi_1-\varphi)  + e^{2\mu_0} \sin^2(\varphi_1-\varphi) \right)\cos^2\theta\right] d\theta^2\\
&+ \Delta^{-\frac{1}{2}}\left[e^{2\mu_2}\left(e^{2\mu_0} \cos^2(\varphi_1-\varphi) + e^{-2\mu_0} \sin^2(\varphi_1-\varphi) \right)\sin^2\theta  \,\mathcal{D}\varphi_1^2+e^{2\mu_1}\cos^2\theta \, \mathcal{D}\varphi_2^2 \right] \\
&+ 2 \Delta^{-\frac{1}{2}}\,e^{2\mu_2}\sinh(2 \mu_0) \sin(2( \varphi_1-\varphi)) \sin\theta\cos\theta \,d\theta \mathcal{D}\varphi_1\,,
\end{aligned}
\end{equation}
with
\begin{equation}\label{eq:3Dmetrictrunc}
ds^2_3=-\Omega_1^2 \left( d\tau + \frac{k}{1-\xi^2} d\psi\right)^2+ \frac{\Omega_0^2}{(1-\xi^2)^2} (d\xi^2 + \xi^2 d\psi^2) \,,
\end{equation}
\begin{equation}
\mathcal{D}\varphi_i= d\varphi_i + A^{\varphi_i}\quad i=1,2\,,
\end{equation}
\begin{equation}\label{eq:phiC}
e^{2\Phi}=\frac{Q_1}{4 Q_5} \frac{(2 \Delta+X^2)^2}{\Delta}\,,\quad C_0 = \sqrt{\frac{2 Q_5}{Q_1}}\, \frac{X}{2 \Delta+X^2}\,,
\end{equation}
\begin{equation}\label{eq:Delta}
\Delta= e^{2\mu_1}\left(e^{2\mu_0} \cos^2(\varphi_1-\varphi)  + e^{-2\mu_0} \sin^2(\varphi_1-\varphi) \right)\sin^2\theta+e^{2\mu_2}\cos^2\theta\,,
\end{equation}
\begin{equation}\label{eq:Xphi}
X= \sqrt{1-\xi^2}\, \nu \,\cos(\varphi_1-\varphi) \sin\theta\,.
\end{equation}
The 10D uplift is explained in Appendix \ref{app:uplift10}. The 3D data, $\nu$, $\mu_0$, $\mu_1$, $\mu_2$, $\Omega_0$, $\Omega_1$, $k$ and the components of $A^{\varphi_i}$, could in general be functions of $\xi$, $\tau$ and $\psi$, but for all the solutions described in this article they will just be functions of $\xi$: this is the essence of the ``Q-ball trick" used in \cite{Ganchev:2021pgs}. We will moreover work in a gauge where the $\xi$ components of the gauge fields $A^{\varphi_i}$ vanish. The phase $\varphi$ will have the generic form 
\begin{equation}
\varphi=\omega \tau+ n \psi\,,
\end{equation}
with $\omega$ a real positive number and $n$ an integer: as we will see, in the regime when the solution is a linear perturbation around AdS$_3\times S^3$, $\omega$ determines the twist (i.e. the energy minus the momentum) of the perturbation and $n$ its momentum. 

To characterise the degrees of freedom encoded in the truncation \eqref{eq:metric6D3D}, we look at linear deformations around the AdS$_3\times S^3$ solution, describing the vacuum of the dual CFT, which is obtained for
\begin{equation}
\nu=\mu_0=\mu_1=\mu_2=0\,,\quad A^{\varphi_1}= A^{\varphi_2}=0\,,\quad \Omega_0=\Omega_1=1\,,\quad k=\xi^2\,.
\label{eq:ads_vacuum}
\end{equation}
As it was already noted in \cite{Ganchev:2021pgs}, the only fields whose linearised equations admit regular and normalisable solutions, and that one can thus freely turn on, are $\nu$ and $\mu_0$. Such perturbations are dual to operators of the dual D1-D5 CFT that we would like to identify. 

Regular and normalisable solutions for $\nu$ exist for $\omega = 1+2m$ with $m$ a non-negative integer and are 
\begin{equation}\label{eq:nu}
\nu=\alpha \,\xi^n \, {}_2F_1(-m,m+n+1,n+1,\xi^2)\,,
\end{equation} 
where the hypergeometric function is a polynomial of order $m$ in $\xi^2$ and we have adopted the notation $\alpha_1=\alpha$, controlling the fraction of constituents of that species, which will be identified in \eqref{eq:LLO12}. Note that $\alpha$ sets the scale of the perturbation. At linear order in $\alpha$ the only non-trivial fields are a 3-form $H_{(3)} = d B_{(2)}$ and the axion $C_{(0)}$ which are given by
\begin{equation}\label{eq:C0pert}
\begin{aligned}
C_{(0)} &=\alpha \,\sqrt{\frac{Q_5}{2 Q_1}}\,\mathrm{Re} (B_1 Y_1)\,,\\
H_{(3)} &= \alpha\,\sqrt{\frac{Q_1 Q_5}{2}}\,\mathrm{Re} \left[d(\star_{AdS_3} d B_1 Y_1 -B_1 \star_{S^3} dY_1)\right]\,,
\end{aligned}
\end{equation}
with
\begin{equation}
B_1 = \sqrt{1-\xi^2} \xi^n \, {}_2F_1(-m,m+n+1,n+1,\xi^2)\,e^{-i(2m+1)\tau-in \psi}\,\,,\,\,Y_1 = \sin\theta \,e^{i \varphi_1},
\end{equation}
and $\star_{AdS_3}$, $\star_{S^3}$ the Hodge duals with respect to the undeformed AdS$_3$ and $S^3$ metrics.
$B_1$ and $Y_1$ are scalar harmonics of AdS$_3$ and $S_3$, respectively: $B_1$ is an eigenfunction of $L_0$ and $\tilde L_0$ with eigenvalues
\begin{equation}
h = m+n+\frac{1}{2}\,,\quad \bar h=m+\frac{1}{2}\,,
\end{equation}
and $Y_1$ is an eigenfunction of $J^3_0$ and $\tilde J^3_0$ with eigenvalues
\begin{equation}
j = \bar j=\frac{1}{2}\,.
\end{equation}
For $m=n=0$ the perturbation \eqref{eq:C0pert} is thus dual to a CPO that we denote by $O_{\frac{1}{2},\frac{1}{2}}$ \cite{Kanitscheider:2007wq}. At the CFT orbifold point, $O_{\frac{1}{2},\frac{1}{2}}$  is identified with a bi-linear of the CFT elementary fermions (see for instance \cite{Avery:2010qw,Giusto:2015dfa} for an introduction to the orbifold CFT and the notation we use)
\begin{equation}
O_{\frac{1}{2},\frac{1}{2}} = \frac{\epsilon_{AB}}{\sqrt{2}}\, \psi^{+A}\tilde \psi^{+B}\quad (A,B=1,2)\,.
\end{equation}
From the supergravity perspective, this state is the lowest KK mode in the $S^3$ reduction of one of the tensor multiplets of the 6D theory. For non-vanishing values of $m,n$ the field \eqref{eq:C0pert} is a Virasoro descendant of the CPO given by\footnote{As usual, $|O\rangle$ indicates the state related to the operator $O$ under the standard CFT operator-state correspondence.}
\begin{equation}\label{eq:LLO12}
L_{-1}^{n+m} \tilde L_{-1}^m \,|O_{\frac{1}{2},\frac{1}{2}}\rangle \,.
\end{equation}
A similar, but slightly more complicated, description applies to $\mu_0$. Regular and normalisable solutions for $\mu_0$ exist for $\omega = 1+2m$ where this time $m$ can be a non-negative integer or half-integer, and are proportional to
\begin{equation}\label{eq:mu0}
\mu_0=\beta\,\xi^{2n} (1-\xi^2) \, {}_2F_1(-2m,2m+2n+2,2n+1,\xi^2)\,,
\end{equation} 
where again the hypergeometric function is a polynomial of order $m$ in $\xi^2$, and this time we take $\alpha_2=\beta$. At first order in $\beta$ the non-trivial fields of the truncation are the metric, the dilaton, and the three-form field strength $F_{(3)} = d C_{(2)} - C_{(0)} \, dB_{(2)}$:
\begin{equation}\label{eq:mu0pert}
\begin{aligned}
\frac{ds^2_6}{R_{AdS}^2}&= ds^2_{AdS_3}+ds^2_{S^3}
\\
&+\beta \,\mathrm{Re}\left[B_2 Y_2 \left( ds^2_{AdS_3}+ds^2_{S^3} -2 \frac{d\theta^2}{\sin^2\theta} + 2 \cos^2\theta (d\varphi_1^2-d\varphi_2^2) - 4 i \frac{\cos\theta}{\sin\theta} d\theta d\varphi_1\right) \right]
\\
&= ds^2_{AdS_3}+ds^2_{S^3}+\beta \,\mathrm{Re}\left[B_2 Y_2 \left(ds^2_{AdS_3}-3 ds^2_{S^3}\right) -B_2 \overline{\nabla}_\alpha \partial_\beta Y_2 \,dx^\alpha dx^\beta)\right]\,,\\
e^{2\Phi}&=\frac{Q_1}{Q_5} \left[1+2 \beta\,\mathrm{Re}\left(B_2 Y_2\right)\right]\,,
\\
F_{(3)} &= 2 Q_5 \,\qty( - \text{vol}_{AdS_3} \,+\, \text{vol}_{S^3} )  \, +\, \beta \,Q_5 \,\Re\!\big[\,d \qty(B_2 \star_{S^3} d Y_2)\,\big]\,,
\end{aligned}
\end{equation}
where $ds^2_{AdS_3}$, $ds^2_{S^3}$ are the unperturbed metrics~\eqref{eq:ads3s3m}, $\alpha,\beta$ ($\mu,\nu$) are $S^3$ (AdS$_3$) indices, the covariant derivative $\overline{\nabla}$ is computed with the unperturbed metric. The functions 
\begin{equation}
B_2 = \xi^{2n} (1-\xi^2) \, {}_2F_1(-2m,2m+2n+2,2n+1,\xi^2)\,e^{-2i(2m+1)\tau-2 i n \psi}\,\,,\,\,Y_2 = \sin^2\theta \,e^{2 i \varphi_1}
\end{equation}
are AdS$_3$ and $S^3$ harmonics with eigenvalues
\begin{equation}\label{eq:hhjj11}
h = 2m+2n+1\,,\quad \bar h=2m+1\,,\quad j = \bar j=1\,.
\end{equation}
In the CPO limit $m=n=0$, one finds an operator of dimension $(1,1)$ that we denote by $O_{1,1}$, and for non-vanishing $m$, $n$ we have again its Virasoro descendants
\begin{equation}\label{eq:LLO11}
L_{-1}^{2(n+m)} \tilde L_{-1}^{2m }\,|O_{1,1}\rangle \,.
\end{equation}
The precise identification of the CPO $O_{1,1}$ is more involved: in the orbifold CFT there are two operators with the required quantum numbers, which are the operator built out of the four fermions, $\psi^{+1}\psi^{+2}\tilde \psi^{+1}\tilde \psi^{+2}$, and the supersymmetric twist field of order 3, $\Sigma^{++}_3$. Linear combinations of these two CPOs \cite{Taylor:2007hs,Giusto:2019qig,Rawash:2021pik} form the KK modes of dimension two of a  6D tensor multiplet (different from the one associated with $O_{\frac{1}{2},\frac{1}{2}}$) and of the 6D gravity multiplet; following the notation of~\cite{Rawash:2021pik}, we denote these CPOs as $s_2$ and $\tilde\sigma_2$ respectively. The operator $s_2$ is dual to a perturbation of AdS$_3\times S^3$ \cite{Deger:1998nm} where only the dilaton and the anti-delf-dual part of the RR 3-form are excited, while $\tilde\sigma_2$ only perturbs the 6D metric. The linearised solution \eqref{eq:mu0pert} contains both types of perturbations and thus we conclude that the CPO $O_{1,1}$ associated with $\mu_0$ is a linear combination of the tensor and of the gravity multiplet CPOs:
\begin{equation}\label{eq:O11mu0}
O_{1,1}\sim \Big( \sqrt{3} \,s^{(1,1)}_2 - \,\tilde\sigma^{(1,1)}_2\Big) +  \Big( \sqrt{3} \,s^{(-1,-1)}_2 - \,\tilde\sigma^{(-1,-1)}_2\Big)\,,
\end{equation}
where we use the notation of~\cite{Rawash:2021pik}. The relative coefficient between $s_2$ and $\tilde{\sigma}_2$ is determined in Appendix~\ref{app:3d-CFT} by comparison with a simple BPS solution, while the sign between the contributions of spin $j=\bar{j}=1$ and  $j=\bar{j}=-1$ corresponds to the presence of $\mathrm{Re}\left[B_2 Y_2\right]$ in~\eqref{eq:mu0pert}. The orthogonal linear combination does not have a normalisable deformation within the truncation used here, but is partially encoded in the scalars $\mu_{1,2}$. As we will see in explicit solutions, their presence is necessary for regularity since these fields can be produced at quadratic order when the deformations~\eqref{eq:nu} and~\eqref{eq:mu0} are present.

We would also like to provide an explicit link between the metric perturbation as written in \eqref{eq:mu0pert} and the form used in \cite{Deger:1998nm,Shigemori:2013lta}. For these purpose it is useful to perform the diffeomorphism generated by the 6D vector
\begin{equation}
  \label{eq:zetavec}
  \frac{\beta}{2} \mathrm{Re}\left[\left(B_2 \partial_\alpha Y_2 dx^\alpha - \partial_\mu B_2 Y_2 dx^\mu\right)\right]\;,
\end{equation}
which transforms the metric deformation in eq.~\eqref{eq:mu0pert} into 
\begin{equation}
  \label{eq:defcovf}
  \frac{ds^2_6}{R_{AdS}^2} = ds^2_{AdS_3}+ds^2_{S^3} + \beta\, \mathrm{Re}\left[B_2 Y_2 \left(ds^2_{AdS_3} - 3 ds^2_{S^3}\right) - Y_2 \overline{\nabla}_\mu \partial_\nu B_2 \,dx^\mu dx^\nu  \right],
\end{equation}
that sits in the gauge used in \cite{Deger:1998nm}; in particular, for the CPO with $m=n=0$, the metric \eqref{eq:defcovf} coincides with the one given in eq. (4.5) of \cite{Shigemori:2013lta}. 

As a final remark, from the CFT point of view, it is of course possible to consider half integer values of $n$ in eq.~\eqref{eq:LLO11}. However, the truncation used here and the ``Q-ball trick'' of~\cite{Ganchev:2021pgs} (as mentioned after eq.~\eqref{eq:Xphi}) link the deformations~\eqref{eq:nu} and~\eqref{eq:mu0}. Thus if both of them are non-trivial ($\alpha,\beta\not=0$) then $n$ must be integer in this approach so as to ensure that $C_0$ in~\eqref{eq:C0pert} has the expected periodicity for $\psi \to \psi +2\pi$. If $\alpha=0$, 
all non-trivial fields are periodic functions of $2\varphi$ and it is possible to describe within this formalism also the supergravity dual of the states~\eqref{eq:LLO11} with half-integer $n$. General two-mode solutions, dual to a multiparticle states involving both~\eqref{eq:LLO12} and~\eqref{eq:LLO11} with different values of $n$, fall outside the framework discussed here.

\section{Asymptotically AdS$_3$ solutions}
\label{sec:sugra}

As shown in~\cite{Mayerson:2020tcl,Houppe:2020oqp,Ganchev:2021pgs}, the 3D gauge supergravity briefly summarised in the previous section is rich enough to capture a class of superstrata, see~\cite{Shigemori:2020yuo} for a review. These are regular solutions of type IIB supergravity which admit an AdS$_3 \times S^3$ near-horizon limit and a holographic interpretation in terms of heavy states in the D1-D5 CFT, {\rm i.e.} states whose conformal dimension is proportional to the central charge $c=6 N$. These states can be thought of as multiparticle operators whose constituents are the standard single-particle supergravity modes. The key point is that when the number of such constituents is large ({\rm i.e} a finite fraction of $N$), then the state is massive enough to produce a classical non-trivial backreaction on the bulk geometry deforming it from the vacuum AdS$_3 \times S^3$.

As a warming up, we review in section~\ref{sec:bpsstr} how to fit in the 3D description the BPS state with $N_1$ copies of the single-particle constituent \eqref{eq:LLO12} with $m=0$. This can be easily done keeping the ratio $N_1/N$ exact, thus fitting in the consistent truncation the full non-linear solution \cite{Ganchev:2021pgs}. Then we consider the non-BPS case focusing on states whose constituents are still supergravity modes, but carry, in the CFT language, both left and right moving momentum. In section~\ref{sec:nonbpsalpha} we first focus on the non-BPS version of the BPS superstratum mentioned just above, which means that we consider the same heavy state but now made out of single-particle constituents \eqref{eq:LLO12} with $m\not=0$. By following~\cite{Ganchev:2021pgs} we simplify the non-BPS analysis by working perturbatively in the parameter $N_1/N$. In the supergravity solution this parameter is encoded in the coefficient $\alpha$ of the normalisable deformation~\eqref{eq:nu}, where, as we will see, $\alpha^2$ is related to $N_1/N$. At the linear level in $\alpha$ the non-BPS solution with $m\not=0$ is related with the BPS one by the action of the right-moving Virasoro generator $\tilde L_{-1}$. The novelty for $m\not=0$ appears  at the non-linear level, where regularity requires to switch on the other normalisable deformation available in the truncation: in practice one has to introduce also the deformation~\eqref{eq:mu0} where the parameter $\beta$ is fixed to be proportional to $\alpha^2$ to ensure regularity. In the context of microstate geometries, an analogous mechanism is a general feature of BPS solutions as well~\cite{Giusto:2013bda,Bena:2015bea,Bena:2017xbt} and is known as coiffuring~\cite{Bena:2013ora,Bena:2014rea}.

Of course it is possible to consider the deformation~\eqref{eq:mu0} by itself -- so that at linear order one has $\alpha=0$ and $\beta\not=0$ -- and we show in section \ref{sec:nonbpsbeta} that also in this case there exists a regular perturbative non-BPS solution whose dual heavy state is constructed from single-particle constituents given by~\eqref{eq:LLO11} with $m\not=0$. However, it is not possible to turn on both deformations \eqref{eq:LLO12} and \eqref{eq:LLO11} at the same time, as there is no regular and asymptotically AdS solutions with both $\alpha, \beta \neq 0$ at linear order. We believe this is a limitation of the ansatz, due to the fact that the frequencies of the two scalars, $\nu$ and $\mu_0$, are locked together.

\subsection{The $(1,0,n)$ BPS superstrata}
\label{sec:bpsstr}

The geometry dual to the condensation of a large number of states \eqref{eq:LLO12} with $m=0$ was derived in~\cite{Bena:2017xbt}, where the parameter controlling the number of elementary constituents of the type \eqref{eq:LLO12} was denoted by $b$. The metric is written in a dimensionally reduced form in~\cite{Giusto:2020mup} -- see in particular eqs.~[4.1] and~[5.3]--[5.6] where the dimensionless radial coordinate $r/a$ there should be identified with $\rho=\xi/\sqrt{1-\xi^2}$ of this paper and the angles $\phi$, $\psi$ there with\footnote{The shifts by $\tau$ and $\sigma$ encode the spectral flow needed to obtain the NSNS sector version of the solutions of~\cite{Bena:2017xbt} which are written in the RR sector.} $\tau-\varphi_1$, $\sigma-\varphi_2$ here. By comparing the metric~\eqref{eq:metric6D3D} with the equations mentioned above, we can check that the $(1,0,n)$ BPS superstrata fit in the 3D truncation. This has been first noted in \cite{Mayerson:2020tcl}: we recall the same analysis here both to be self-contained and to spell out the gauge choices we adopt in this article. 

The scalars can be derived by comparing the explicit result for the axio-dilaton of the superstrata~\cite{Bena:2017xbt} with~\eqref{eq:phiC}, from which we find
\begin{equation}\label{eq:scalars10n}
\begin{aligned}
\Delta &=\left[1-\frac{b^2 R_y^2}{2 Q_1 Q_5}(1-\xi^2) \xi^{2n}\right] \sin^2\theta+\cos^2\theta\,,\\
X&=-\sqrt{2} \frac{b R_y}{\sqrt{Q_1 Q_5}}\sqrt{1-\xi^2} \xi^n\sin\theta \cos(\tau-\varphi_1+ n \psi)\,.
\end{aligned}
\end{equation}
We can now match these expressions with \eqref{eq:Delta}, \eqref{eq:Xphi}: from the fact that $\Delta$ does not depend on $\varphi_1$ we immediately infer the vanishing of $\mu_0$ and from the coefficients of $\sin^2\theta$ and $\cos^2\theta$ we read off $\mu_1$ and $\mu_2$:
\begin{equation}\label{eq:mu10n}
e^{2\mu_1} = 1-\frac{b^2 R_y^2}{2 Q_1 Q_5} (1-\xi^2)\xi^{2n} \,\,, \quad \mu_0=\mu_2=0\,.
\end{equation}
We note that the phase of $X$ is consistent with $\omega=1$ for $m=0$ and its modulus gives
\begin{equation}
\nu =-\sqrt{2} \frac{b R_y}{\sqrt{Q_1 Q_5}} \xi^n\;.
\end{equation}
In the notation of Section \ref{sec:trunCFT}, the strength of the perturbation  \eqref{eq:LLO12} is
\begin{equation}
  \alpha = -\sqrt{2} \frac{b R_y}{\sqrt{Q_1 Q_5}} \,.
\end{equation}

One can now check that the above values of $\mu_i$ reproduce, via \eqref{eq:metric6D3D}, the $S^3$-part of the 6D metric as given in eq. (5.3) of~\cite{Giusto:2020mup}. Then one considers the vectors given in eq.~[5.4] of~\cite{Giusto:2020mup}: after taking into account the effect of the spectral flow we read the following gauge fields
\begin{equation}
  \label{eq:Astra}
A^{\varphi_1} = -\frac{b^2 R_y^2}{2 Q_1 Q_5}  d\tau \;, \quad A^{\varphi_2} = \frac{b^2 R_y^2}{2 Q_1 Q_5} \frac{d\tau - (1-\xi^2) \xi^{2n} \left(d\tau+ \frac{\xi^2}{1-\xi^2}d\psi\right)}{1-\frac{b^2 R_y^2}{2 Q_1 Q_5} (1-\xi^2) \xi^{2n}}\,.
\end{equation}
One can remark that the $\sigma$ component of both $A^{\varphi_1}$ and $A^{\varphi_2}$ vanishes at $\xi=0$, where the $\sigma$-circle shrinks, while the $\tau$-components of the gauge fields at the asymptotic point $\xi=1$ reproduce the angular momenta of the superstrata, according to the formulas given in Section~\ref{sec:holo}. Finally one extracts $\Omega_0$, $\Omega_1$ and $k$ by comparing eq.~[5.5] of~\cite{Giusto:2020mup} with~\eqref{eq:3Dmetrictrunc}. It is useful to note that the prefactor of $ds^2_3$ in \eqref{eq:metric6D3D}, $e^{-2(\mu_1+\mu_2)}\Delta^{1/2}$, agrees with that appearing in ~\cite{Giusto:2020mup}: $V^{-2}=\frac{(Q_1 Q_5)^{3/2} \sin^2\theta \cos^2\theta}{g_{\theta\theta} g_{\varphi_1\varphi_1}g_{\varphi_2\varphi_2}}$. Then one finds
\begin{equation}
\begin{aligned}
\Omega_0^2&=\frac{g_{\xi\xi}}{R_{AdS}^2} (1-\xi^2)^2=\frac{g_{rr}}{R_{AdS}^2} a^2(1-\xi^2)^{-1} = 1-\frac{b^2 R_y^2}{2 Q_1 Q_5}(1-\xi^2)\xi^{2n} \,,\\
\Omega_1^2&=-\frac{g_{\tau\tau}}{R_{AdS}^2}=\frac{a^4 R_y^4}{(Q_1 Q_5)^2}\,,\quad k = - \frac{g_{\tau\psi}}{R_{AdS}^2\Omega_1^2}(1-\xi^2)= \frac{Q_1 Q_5}{a^2 R_y^2}\,\xi^2\,,
\end{aligned}
\end{equation}
where 
\begin{equation}
a^2+\frac{b^2}{2}=\frac{Q_1 Q_5}{R_y^2}= a_0^2\,.
\end{equation}
Note that $\Omega_1$ and $k^2/\xi^2$ are constants different than one for $b\not=0$. The values of these constants are required so that 
\begin{equation}
\lim_{\xi\to 1} \frac{g_{\sigma\sigma}}{g_{\tau\tau}}=-1\,;
\label{eq:time_condition}
\end{equation}
this gauge condition defines a canonically normalised time $\tau$, and it will be imposed throughout this article.

We conclude this subsection by discussing the CFT interpretation of the absence of the field $\mu_0$ in the BPS solution (see also Appendix~\ref{app:3d-CFT} for some further detail). We know that its normalisable modes~\eqref{eq:mu0} correspond to the expectation value of a CPO of dimension $h=\bar{h}=1$ and its AdS descendants. Since this property does not depend on the value of $n$, we can focus on the simplest case ($n=0$) and use eq.~(D.4) of \cite{Rawash:2021pik} to derive, from the asymptotic behaviour of the solution, the expectation values of the relevant CPOs of such dimension ({\rm i.e.} $s_2$ and $\tilde\sigma_2$). As expected, the only non-trivial components have\footnote
{Notice that this is the result of a non-trivial cancellation in the case of $\tilde\sigma_2$.} $j=\bar{j}=1$, consistently with~eq.~\eqref{eq:hhjj11}. From these explicit results, one can see that the combination in~\eqref{eq:O11mu0} has vanishing expectation value in the $n=0$ superstratum and so should be identified with $\mu_0$\footnote{Note that this remark refers exclusively to the $(1,0,n)$ superstrata. New supersymmetric solutions fitting the Q-ball ansatz have been discovered for which $\mu_0 \neq 0$, see \cite{Ganchev:2021new}}. The orthogonal combination has, instead, a non-trivial expectation value which is encoded in the field $\mu_1$. In the construction of \cite{Bena:2015bea,Bena:2016ypk}, this non-trivial expectation value followed from a coiffuring mechanism which required adding to one of the warp factors, $Z_1$, an appropriate term of order $b^2$, as allowed by the linear structure of the BPS equations. Note, however, that in the 3D truncation the field $\mu_1$ is uniquely determined by the field equations after specifying the value of $\nu$ and thus, the coiffuring mechanism of \cite{Bena:2015bea,Bena:2016ypk} is automatically implied by the truncation.



\subsection{Perturbative non-BPS solutions} 
\label{sub:gauge_fixing_the_solutions}

We now turn to the non-BPS solutions associated to the deformations (\ref{eq:LLO12}) and (\ref{eq:LLO11}), when $m \neq 0$. To simplify the analysis, we are focusing on the case $m=1$. While the exact solution corresponding to a finite deformation seems out of reach, one can learn more about these solutions by extending to higher orders the perturbative expansion around AdS$_3\times S^3$ which was started in section \ref{sec:trunCFT}. Moreover, in this work, we will only consider the deformations turned on one at a time, since the solutions involving both deformations do not fit our ansatz; we have verified this fact by an explicit supergravity analysis and will provide an intuitive explanation based on a CFT perspective in Section~\ref{sec:holo}.

Our perturbative expansion has the following form:
\begin{equation}
\nu ~=~ \nu^{AdS} \,+\, \epsilon \,\delta^{(1)} \nu \,+\, \epsilon^2 \,\delta^{(2)} \nu \,+\, \dots \,,
\label{eq:perturbExp}
\end{equation}
where $\epsilon$ will be either $\alpha$ or $\beta$,
and we preform a similar expansion for all the other fields, as well as for the frequency. The superscript $AdS$ denotes the AdS vacuum, given by (\ref{eq:ads_vacuum}). Furthermore, we will sometimes work with a slightly different basis of fields that makes the computations easier, namely:
\begin{equation}
\mu_{\pm} = \frac{1}{2}\big(\mu_1\pm\mu_2\big),\quad
\quad\Phi_{\pm} = \frac12 (A^{\varphi_1}_\tau \pm \,A^{\varphi_2}_\tau),
\quad\Psi_{\pm} = \frac12 (A^{\varphi_1}_\psi \pm \,A^{\varphi_2}_\psi).
\label{eq:basis_fields}
\end{equation}

We can then solve the system of second order, ordinary differential equations, order by order in $\epsilon$. The linearised equations of motion are given in \cite{Ganchev:2021pgs} and recalled in Appendix \ref{app:linearized_oem}. Note that we use the $R_{\tau\psi}$ component of the Einstein equation to obtain a first order equation for $k$, instead of the second order one that results from the $R_{\psi\psi}$ component. At higher orders, the equations one needs to solve are given by the linear piece plus additional sources coming from the previous orders.

Before delving into the computation of the non-BPS solutions, we discuss the set of conditions we need to provide to fully specify the solutions at each order. There are three types of conditions: regularity constraints, normalizability constraints or gauge fixing.

The first conditions arise from the fact that we are only looking for smooth solutions. This implies in particular that all the fields must be regular at the origin. At each order in the perturbative expansion, most of the fields admit solutions containing divergences at the origin, these solutions will have to be discarded by choosing appropriately the integration constants.

Since the $\sigma$-circle shrinks to zero size at the origin, regularity also implies that the components of the gauge fields in this direction must vanish:
\begin{equation}
  A^{\varphi_1}_\sigma(0)=0 \qand A^{\varphi_2}_\sigma(0)=0.
\end{equation}

We will also require that the metric is free of conical deficits at the origin, which gives a condition on $k$, namely,
\begin{equation}
k ~=~ \mathcal{O}(\xi^2) \qq{as} \xi \to 0.
\label{eq:kNoConical}
\end{equation}

In addition to smoothness, we look for solutions that are normalizable. In this context, it means that we require the metric to be asymptotically AdS (with flat boundary), and that each field decays at infinity as fast, or faster, than its normalizable perturbation. For the scalars, this implies
\begin{equation}
\mu_0,\,\mu_-=\mathcal{O}(1-\xi^2),\quad \mu_+=\mathcal{O}\big((1-\xi^2)^2\big),\qand \nu = \mathcal{O}(1) \qq{as} \xi \to 1 \,.
\end{equation}
For the gauge fields, it translates to asking that they limit to a finite value at the boundary. As for the metric functions, we need to impose
\begin{equation}
  \Omega_0(1) = 1 \qand \Omega_1(1) = \frac{1}{k(1)} \,.
\end{equation}
These two conditions are actually already a consequence of the equations of motion.

We now turn to the gauge fixing. The conditions listed above are not sufficient to fully specify a solution, as the ansatz contains residual gauge invariances. First, time can be freely rescaled. As stated in the previous section we will require \eqref{eq:time_condition} to fix this gauge freedom. Second, we can freely shift $A^{\varphi_i}_\tau$ by redefining the angular coordinates, without affecting the 3-dimensional supergravity solution. More precisely, the theory is invariant under the global $U(1)$ transformations
\begin{equation}
  A^{\varphi_1}_\tau \to A^{\varphi_1}_\tau + \theta,\quad\omega \to \omega + \theta\,\qand A^{\varphi_2}_\tau \to A^{\varphi_2}_\tau + \theta',
\end{equation}
where $\theta$ and $\theta'$ are real parameters.

There is no canonical way to fix these symmetries in the bulk. We will thus make the following arbitrary choice
\begin{equation}
  \omega = 2m+1 \qand A^{\varphi_2}_\tau(0) = 0\,,
  \label{eq:gaugeOmega}
\end{equation}
where the first condition effectively means that when performing the perturbative expansion around the AdS$_3$ vacuum, we will impose at each order that the correction to the frequency vanishes: $\delta^{(k)} \omega = 0$ for $k\geq 1$.

However, as we will see in section \ref{sec:holo}, these $U(1)$ symmetries are large gauge transformations, which are anomalous in the CFT, and thus change the state of the field theory. In other words, each state is associated to a fixed asymptotic value of the gauge fields. This means that when matching the gravitational solutions to CFT states, we will have to perform an additional gauge transformation, so that, with the help of the holographic dictionary, we set the gauge fields at infinity to the values of the CFT state whose dual we are constructing.

Finally, we need one last condition to completely fix a solution, and it is the scale of the deformation. For the first class of deformations, when $\epsilon = \alpha$, we do this by fixing $\alpha$, such that, $\nu \sim \alpha \, \xi^n$, at the origin. For the second class of deformation, $\epsilon = \beta$, we choose $\mu_0 \sim \beta \, \xi^{2n}$ at the origin.

Note that our system of linearised equations is overconstrained. Namely, after imposing the necessary 21 conditions, detailed in the text above (10 second-order equations plus 1 first order equation), a solution to the equations exists only if the frequency $\omega$ is quantised, as already implemented at linear order in Section (\ref{sec:trunCFT}). The fact that such solutions exist is non-trivial, and one of the successes of this ansatz.

\subsection{The \texorpdfstring{$\alpha$}{alpha}-class of perturbative non-BPS solutions}
\label{sec:nonbpsalpha}

We will first focus solely on the deformation (\ref{eq:LLO12}), taking $\epsilon=\alpha$ in \eqref{eq:perturbExp}, and effectively setting $\beta=0$ to turn off $\mu_0$ at linear order, see (\ref{eq:mu0}).

It is necessary to compute the expansion up to at least fourth order in $\alpha$, for two reasons. The first is that this order will be needed to perform non-trivial checks in the CFT. The second reason is that the coiffuring condition appears at this order. This non-trivial result shows that, in order for the $\alpha$-deformation to be normalizable, one needs to also excite the scalar $\mu_0$ (at higher than linear order), corresponding to the $\beta$-deformation, in a precise relation to $\alpha$.

\subsubsection*{First order}

The linear perturbation has already been described in section \ref{sec:trunCFT}. Only one scalar field is excited, \eqref{eq:nu} with $m=1$,
\begin{equation}
  \delta^{(1)} \nu ~=~ \xi^n \,\qty(1 - \frac{n+2}{n+1} \xi^2) \,,
\end{equation}
and the other fields are set to zero. Note that $\alpha$ is normalised so that $\nu \sim \alpha\,\xi^n$ as $\xi \to 0$.

\subsubsection*{Second order}

At order $\alpha^2$, all fields but $\nu$ receive normalisable corrections. 
One could argue that the scalar $\nu$, as well as $\mu_0$, have normalizable homogeneous solutions, which can always be added to the two scalars at each order. Indeed, as we will see, the homogeneous term in $\mu_0$ will prove to be very important for assuring the normalisability of the solutions. On the other hand, this term in $\nu$ would only lead to a redefinition of $\alpha$, and therefore can be discarded.  

For the scalars, we find
\begin{align}
  \delta^{(2)} \mu_0 ~&=~ \frac{n+2}{8(2n+1)(n+1)^2} \xi^{2+2n} (1-\xi^2) \qty((2n+3) \xi^2 - 2(n+1)) \nonumber \\
  & \quad + \beta_2 \, \xi^{2n} (1-\xi^2) \qty(1-4 \frac{n+2}{2n+1} \xi^2 + \frac{(n+2)(2n+5)}{(n+1)(2n+1)} \xi^4),
  \\
  \delta^{(2)} \mu_1 ~&=~ - \frac{1}{8} \xi^{2n} (1-\xi^2) \qty(1- \frac{n+2}{n+1}\xi^2)^2,
  \\
  \delta^{(2)} \mu_2 ~&=~ 0 \,,
\end{align}
where we named $\beta_2$ the constant of integration for the homogeneous solution in $\mu_0$. We will find that it gets fixed at order $\alpha^4$ by a coiffuring mechanism.

The corrections to the gauge fields are
\begin{align}
   \delta^{(2)} A^{\varphi_1} ~&=~ \gamma_2 \, d\tau, \\
   \delta^{(2)} A^{\varphi_2} ~&=~ \frac1{4(n+1)^2} \xi^{2n} \Big[\qty(1+n(n+2)(1-\xi^2)^2) d\tau \nonumber\\
   & \qquad\qquad\qquad\qquad + \xi^2 \qty((n+1)(n+3) (1-\xi^2)^2 + \xi^4) d\sigma \Big],
 \end{align}
 where $\gamma_2$ is a constant, which will be fixed at the next order by demanding regularity of $\nu$ at infinity and imposing the gauge fixing \eqref{eq:gaugeOmega}.  Again, the subscript denotes the order at which these constants are introduced.

 The corrections to the metric are
 \begin{align}
   \delta^{(2)} \Omega_0 ~&=~ - \frac1{8(n+1)^2} \xi^{2n} (1-\xi^2) \qty[(n+1)^2 - 2 (n^2 + 3n + 1) \xi^2 + (n+2)^2 \xi^4],
   \\
   \delta^{(2)} \Omega_1 ~&=~ - \frac3{4(n+1)^2} + \frac1{2(n+1)^2} \xi^{2n+2} (1-\xi^2),
   \\
   \delta^{(2)} k ~&=~  \frac3{4(n+1)^2} \xi^2 + \frac1{2(n+1)^2} \xi^{2n+2} (1-\xi^2) \qty(n+1 -(n+2)\xi^2).
\end{align}

The factor of $3/4$ appearing in $\Omega_1$ and $k$ is a result of the normalisation of time (\ref{eq:time_condition}), it is different from the choice made in \cite{Ganchev:2021pgs}, where the gauge was fixed by requiring $\Omega_1(1) = 1$.

\subsubsection*{Third order}

As we progress to higher orders, it becomes harder to solve the equations for a general mode number $n$, as the source terms become more complicated. Two methods can be used to determine the solutions. The first is based on the fact that all normalizable solutions appear to be polynomials in $\xi$. The differential equations thus become recurrence equations, that can sometimes be solved. The second method is to solve the equations for a large number of different values of $n$, and to try to extrapolate the general solution. The result can then be checked against the original equations. We use the latter method.

Akin to the linear order situation, at third order in $\alpha$ only $\delta^{(3)}\nu$ receives a non-trivial correction (non-linearly sourced by lower order solutions). We again find a non-normalisable (log-divergent at infinity) piece, whose vanishing gives us an expression for the gauge invariant quantity $\delta^{(2)}\omega-\delta^{(2)}A^{\varphi_1}_\tau$. We then set $\delta^{(2)}\omega$ to zero in order to fix $A^{\varphi_1}_\tau$ at infinity, as explained in the previous section. This results in:
\begin{equation}
\gamma_2 ~=~ \frac{9}{4(n+1)^2} - \frac3{32(n+1)^2} \qty(\frac1{2n+1} + \frac{22}{2n+3} + \frac{1}{2n+5}) \,.
\end{equation}

Moreover, one can add homogeneous solutions\footnote{Solutions to the homogeneous equations in Appendix \ref{app:linearized_oem}} in $\delta^{(3)}\mu_0$ and $\delta^{(3)} A^{\varphi_1}_\tau$, by introducing respectively the constants $\beta_3$ and $\gamma_3$. We have:
\begin{align}
\delta^{(3)}\nu ~=&~  \frac{3 (3n^2+9\,n+4) \, \xi^{2+n}}{2(8n^3+36n^2+46n+15)(n+1)^3} \qty[\sum_{k=1}^{n-1} \frac{n+k+2}{k(k+1)} \xi^{2k}-(n+3)] -\xi^{3n} \mathcal{P}_n(\xi^2)\,,
\\
\delta^{(3)}\mu_0 ~=&~ \beta_3\,\xi^{2n} (1-\xi^2) \qty(1-2 \frac{2n+4}{2n+1} \xi^2 + \frac{(n+2)(2n+5)}{(n+1)(2n+1)} \xi^4) \,,
\\
\delta^{(3)} A^{\varphi_1}_\tau ~=&~ \gamma_3 \,,
\end{align}
where $\mathcal{P}_n(\xi^2)$ is a family of polynomials of order $4$ in $\xi^2$, whose terms are complicated rational functions of the mode $n$, and have a linear dependence on $\beta_2$.

The constant $\gamma_3$ will be fixed at next order, by the regularity of the scalars at infinity and $\delta^{(3)}\omega=0$, and $\beta_3$ will get fixed two orders later (at $\alpha^5$) to avoid non-normalisable solutions, as explained earlier.

\subsubsection*{Fourth and higher orders}

At fourth order in the perturbative expansion, one finds that all fields but $\nu$ receive corrections, as was the case at order $\alpha^2$. Only the scalar $\delta^{(4)} \mu_0$ has been computed for a general value of $n$. We find that it is normalizable (and polynomial) if and only if the following coiffuring constraint is satisfied
\begin{equation}
  \beta_2 ~=~ -\frac{1}8
  \qand \gamma_3 ~=~ 0\,.
\end{equation}

At higher orders, while the analysis becomes too involved when keeping $n$ general, it is possible to continue it at a fixed value of $n$. We did this up to tenth order, and for $n$ up to $10$. We find patterns that we postulate are true for all modes, in particular

\begin{align}
\delta^{(2\,k)}\nu=0,\qand\delta^{(2\,k+1)}\mu_0=0 \ ,&\qfor k\in\mathbb{N} \,, \\
\gamma_k ~=~ 0 \qand \beta_k ~=~ 0 \ , &\qfor k \geq 3 \,.
\end{align}

\subsection{The \texorpdfstring{$\beta$}{beta}-class of perturbative non-BPS solutions}
\label{sec:nonbpsbeta}

We now repeat the previous analysis with the solutions emerging from the second deformation (\ref{eq:LLO11}). That is, we take $\epsilon=\beta$, turn off the $\nu$ deformation by setting $\alpha = 0$, and make a perturbative expansion of the solution in small $\beta$. The $\beta$-deformation does not require a coiffuring. While our results with general $n$ are limited to the second order, we will continue the analysis for a few fixed values of $n$, as done for the $\alpha$-deformation.

\subsubsection*{First order}

The linear solution has been described in section \ref{sec:trunCFT}. The only excited 3D field is
\begin{equation}
  \delta^{(1)}\mu_0 ~=~ \left(1-\xi ^2\right) \xi ^{2 n} \qty(1-2 \frac{2n+4}{2n+1} \xi^2 + \frac{(n+2)(2n+5)}{(n+1)(2n+1)} \xi^4) \,.
\end{equation}

\subsubsection*{Second order}

At second order, most of the fields receive a correction. The scalars are given by
\begin{equation}
  \delta^{(2)} \nu ~=~ \delta^{(2)} \mu_0 ~=~ 0,
\end{equation}
and
\begin{equation}
\begin{aligned}
  \delta^{(2)} \mu_1 ~&=~ -\delta^{(2)} \mu_2 ~=~ \xi ^{4 n+2} \Bigg(\frac{12 (n+2) \left(4 n^2+12 n+7\right) \xi ^4}{(n+1) (2 n+1)^2 (2 n+3)^2}+\frac{2 (n+2)^2 \xi ^8}{(n+1)^2 (2 n+1)^2} \\
  & -\frac{4 (2 n+5) \xi ^6}{(n+1) (2 n+1)^2}-\frac{4 (n+2) \xi ^2}{(n+1)^2 (2 n+1)}+\frac{2}{(2 n+1)^2}\Bigg)-\frac{6}{(n+1)^2 (2 n+1)^2 (2 n+3)^2} \,.
\end{aligned}
\end{equation}
The gauge fields are more appropriately written in the basis  described in (\ref{eq:basis_fields}). They are
\begin{equation}
\begin{aligned}
  \delta^{(2)} \Phi_+ ~&=~ \frac{-6}{(2 n+1)^2} \Bigg[\frac{(n+2)^2 \xi ^8}{(n+1)^2} -\frac{2 (2 n+5) \xi ^6}{n+1} + \frac{6 (n+2) \left(4 n^2+12 n+7\right) \xi ^4}{(n+1) (2 n+3)^2} \\
  &\qquad -\frac{2 (n+2) (2 n+1) \xi ^2}{(n+1)^2}+1\Bigg] \xi ^{4 n+2} + \gamma _2.
\end{aligned}
\end{equation}
\begin{equation}
  \delta^{(2)} \Phi_- ~=~ \frac{-4 \left(1-\xi ^2\right)^2 \left((n^2 + 4n +4) \xi ^4-(2 n^2 + 6n +3) \xi ^2+n^2+2 n+1\right) \xi ^{4 n+2}}{(n+1)^2 (2 n+1)^2} + \gamma _2.
\end{equation}
\begin{equation}
\begin{aligned}
    \delta^{(2)} \Psi_- ~&=~ \xi ^{4 n+2} \Bigg[- \frac{(n+2)^2 \xi ^8}{(n+1)^2 (2 n+1)}  + \frac{2 \left(4 n^3+16 n^2+17 n+4\right) \xi ^6}{(n+1)^2 (2 n+1)^2} \\
    &-\frac{2 \left(12 n^4+60 n^3+99 n^2+59 n+9\right) \xi ^4}{(n+1)^2 (2 n+1)^2 (2 n+3)} + \frac{2 \left(4 n^2+8 n+1\right) \xi ^2}{(n+1) (2 n+1)^2} - \frac{1}{2 n+1}\Bigg].
\end{aligned}
\end{equation}
\begin{equation}
\begin{aligned}
  \delta^{(2)} \Psi_+ ~&=~ \frac{-\xi ^2}{(n+1)^2 (2 n+1)^2 (2 n+3)^2} \Bigg[\xi ^{4 n} \bigg((n+2)^2 (2 n-1) (2 n+3)^2 \xi ^8 
  \\
  &\quad -2 (n+1) (2 n+3)^2 \left(4 n^2+9 n-4\right) \xi ^6
  \\
  &\quad +6 (n+1) \left(8 n^4+40 n^3+58 n^2+14 n-9\right) \xi ^4
  \\
  &\quad -2 (2 n+1) \left(8 n^4+42 n^3+73 n^2+39 n-9\right) \xi ^2
  \\
  &\quad + 8 n^5+44 n^4+94 n^3+97 n^2+48 n-9 \bigg)
  -\frac{18 \left(1-\xi ^{4 n}\right)}{1-\xi ^2}\Bigg].
\end{aligned}
\end{equation}
Finally, the corrections to the metric functions are
\begin{equation}
  \delta^{(2)} \Omega_1 ~=~ \frac{6 \left(2 (2 n+3) \left(\xi ^2-1\right)^2 \xi ^{4 n+2} \left((n+2)^2 \xi ^4-(2 n (n+3)+3) \xi ^2+(n+1)^2\right)-3\right)}{(n+1)^2 (2 n+1)^2 (2 n+3)}.
\end{equation}
\begin{equation}
\begin{aligned}
  \delta^{(2)} \Omega_0 ~&=~ \left(1-\xi ^2\right)^2 \xi ^{4 n} \Big(-\frac{(n+2)^2 (2 n+5)^2 \xi ^8}{2 (n+1)^2 (2 n+1)^2} + \frac{2 (n+2)^2 \left(4 n^2+14 n+5\right) \xi ^6}{(n+1)^2 (2 n+1)^2}\\
  &-\frac{\left(12 n^4+72 n^3+133 n^2+75 n+9\right) \xi ^4}{(n+1)^2 (2 n+1)^2}+\frac{2 \left(4 n^2+10 n-1\right) \xi ^2}{(2 n+1)^2}-\frac{1}{2}\Big).
\end{aligned}
\end{equation}
\begin{equation}
\begin{aligned}
  \delta^{(2)} k ~&=~ -\frac{ \xi ^2}{(n+1)^2 (2 n+1)^2 (2 n+3)} \bigg[\left(\xi ^2-1\right) \xi ^{4 n} \Big(5 (n+2)^2 (2 n+3) (2 n+5) \xi ^8 \\
  &-4 (n+2) (2 n+3) (5 n (2 n+7)+23) \xi ^6+2 \qty(n (n+3) \qty(60 n^2 +180 n+209)+171) \xi ^4\\
  &-4 (n+1) (2 n+3) (10 n^2 + 25 n+8) \xi ^2+5 (n+1)^2 (2 n+1) (2 n+3)\Big)-18\bigg].
\end{aligned}
\end{equation}

One could add a homogeneous term to the scalar $\delta^{(2)}\nu$, but we find at next order that it has to vanish. Here again, $\gamma_2$ is an integration constant that is fixed when requiring that the solution is normalizable at third order

\subsubsection*{Third order}

At this order the only field that receives a non-trivial correction is $\mu_0$:

\begin{multline}
\delta^{(3)}\mu_0=-(1-\xi^2)\xi^{2+2\,n}\Bigg[\frac{48\,n_{1,2}\,B_1}{n_{1,1}^2\,n_{2,1}^3\,n_{2,3}^2\,n_{4,3}\,n_{4,5}\,n_{4,7}\,n_{4,9}}-\frac{B_2\,\xi ^2}{n_{1,1}^3\,n_{2,1}^3\,n_{2,3}\,n_{4,3}\,n_{4,5}\,n_{4,7}\,n_{4,9}}\\
+\frac{\xi^{4\,n}}{3}\Bigg(\frac{4\,n_{8,17}+47}{n_{2,1}^2}-\frac{1}{\xi^2}-\frac{B_3\,\xi^2}{n_{1,1}^2\,n_{2,1}^3}+\frac{B_4\,\xi^4}{n_{1,1}^3\,n_{2,1}^4\,n_{2,3}^2\,n_{4,7}\,n_{4,9}}-\frac{B_5\,\xi^6}{n_{1,1}^4\,n_{2,1}^3\,n_{2,3}^2\,n_{4,9}}\\
+\frac{B_6\,\xi^8}{n_{1,1}^3n_{2,1}^3}-\frac{B_7\,\xi^{10}}{n_{1,1}^3\,n_{2,1}^3}
+\frac{n_{1,2}^3\,B_8\,\xi^{12}}{n_{1,1}^3\,n_{2,1}^3}-\frac{B_9\,\xi^{14}}{n_{1,1}^3\,n_{2,1}^3}\Bigg)\\
+\frac{24\,\big(n_{4,3}\,n_{4,5}\,n_{4,7}\big)^{-1}}{n_{1,1}^3\,n_{2,1}^3\,n_{2,3}^3\,,n_{4,9}}\sum_{k=1}^{2n}\frac{\xi^{2k+2}}{k\,k_{1,1}\,k_{1,2}}\bigg(C_1\,n^5-C_2\,n^4-C_3\,n^3-C_4\,n^2-C_5n\\
-1260 \left(k^3-15 k^2-88k-120\right)+7824 (k+8) n^6+5216 n^7\bigg)\Bigg],
\end{multline}
where,
\begin{align}
n_{a,b}=(a\,n+b),
\end{align}
and
\begin{align}
B_1=\,&652 n^4+4052 n^3+9007 n^2+8472 n+2835,\notag\\
B_2=\,&12 \left(1956 n^5+17884 n^4+62649 n^3+104834 n^2+83493 n+25200\right),\notag\\
B_3=\,&224 n^5+1288 n^4+3080 n^3+3790 n^2+2387 n+715,\notag\\
B_4=\,&57344 n^{11}+881664 n^{10}+6151936 n^9+25715904 n^8+71534256 n^7+138984384 n^6\notag\\
+&192383752n^5+189668868n^4+130493723n^3+59681976n^2+16382109n+2083284,\notag\\
B_5=\,&8960 n^{10}+136640 n^9+940800 n^8+3852240 n^7+10389264 n^6+19283532 n^5\notag\\
+&24941848 n^4+22184701 n^3+12969072 n^2+4487253n+693342,\notag\\
B_6=\,&7168 n^9+110208 n^8+756672 n^7+3045864 n^6+7923636 n^5+13814178 n^4\notag\\
+&16129931 n^3+12142293 n^2+5328426 n+1032120,\notag\\
B_7=\,&224 n^6-2520 n^5-11928 n^4-30426 n^3-44139 n^2-34446 n-11200,\notag\\
B_8=\,&(64 n^3+408 n^2+882 n+655),\quad B_9=2 n^2+9 n+10.
\end{align}
\begin{align}
C_1=\,&8\left(489 k^2+9780 k+39580\right),\quad C_2=\,12\left(31 k^3-2352 k^2-26478 k-72850\right),\notag\\
C_3=\,&2\left(1116 k^3-39597 k^2-334938 k-710137\right),\notag\\
C_4=\,&3\left(1579 k^3-36078 k^2-257419 k-451686\right),\notag\\
C_5=\,&9 \left(463 k^3-8028 k^2-51199 k-77900\right).
\end{align}
Moreover, the constant parts in the $\Phi_\pm$ fields are left undetermined until the next order:
\begin{equation}
\delta^{(3)} \Phi_+=\gamma_3\qand\delta^{(3)} \Phi_-=\gamma_3.
\end{equation}

Analogously to the $\alpha$-class calculation, in the process of solving the equations, we need to discard a non-normalisable mode for the scalar field driving the solution, $\mu_0$. This results in an expression for $\delta^{(2)}\omega$ that we set to $0$, in order to fix $A^{\varphi_1}_\tau$ at infinity. This sets the constant $\gamma_2$ to

\begin{equation}
\gamma_2 ~=~ 6\,\frac{2304n^5 + 16360n^4 + 44880n^3 + 59330n^2 + 37671n + 9135}{(n+1)^2(2n+1)^2(2n+3)^2(4n+3)(4n+5)(4n+7)(4n+9)} \,.
\end{equation}

As before, we could have added a homogeneous term to the scalar $\delta^{(3)}\nu$, but it vanishes at next order.

\subsubsection*{Fourth and higher orders}

Once again, the solution at higher order becomes too complicated to determine for a general mode number, but it is possible to continue the computations with a fixed value of $n$, and deduce some general properties of the solution. This has been done up to eighth order, and with $n$ up to $10$.

We find at all orders a polynomial expression for all the fields. At order $k$, one can introduce a constant $\gamma_k$ in the gauge field $A^{\varphi_1}$, and this constant is fixed at order $k+1$, whereby it is zero for odd values of $k$. Furthermore, we find that $\nu = 0$ at every order we compute. We postulate that this is an exact result, and that it holds for all values of $n$.

\section{Properties of the solution and its holographic interpretation}
\label{sec:holo}
The perturbative supergravity analysis of the previous section has shown the existence of regular and normalizable solutions that tend asymptotically to AdS$_3\times S^3\times \mathcal{M}$ and that have, thus, all the properties required to be dual to states of the D1-D5 CFT. Here we will identify the CFT dual states and perform some checks on the proposed holographic dictionary. In the large $N$ limit, the dynamics is described in terms of generalised free fields that are dual to the CPOs and their descendants. The geometries discussed in this paper are dual to a collection of such operators that forms a bound state when $N$ is large, but finite, {\rm i.e.} when the gravitational interaction is switched on.
Given the CFT interpretation of the linear perturbations proportional to $\alpha$ and $\beta$ discussed in section~\ref{sec:trunCFT}, it is natural to identify the geometry where $\alpha$ ($\beta$) is turned on at linear order with a bound state of the descendants $L_{-1}^{n+m} \tilde L_{-1}^m \,|O_{\frac{1}{2},\frac{1}{2}}\rangle$ ($L_{-1}^{2(n+m)} \tilde L_{-1}^{2m} \,|O_{1,1}\rangle$). In the CFT it is perfectly legitimate to consider states with both type of strands, and these should be dual to solutions where both $\alpha$ and $\beta$ appear as free parameters at linear order: we believe that the fact that we were unable to find such solutions is a limitation of the ``Q-ball" ansatz used in section~\ref{sec:sugra} to simplify the gravity construction and not a fundamental obstruction. We will comment on the qualitative reason for this limitation and on the necessary generalisation, later in this section.

For simplicity, we start our analysis with the solution where only $\alpha$ is turned on and, as done in the supergravity construction, with the lowest-energy non-BPS states for fixed momentum and thus take $m=1$ (or $\omega=3$) in \eqref{eq:LLO12}. The schematic\footnote{More precisely, according to the paradigm introduced in \cite{Skenderis:2006ah}, gravity configurations are dual to ``coherent states'' made by linear combinations of energy eigenstates of the form \eqref{eq:dualstate} peaked over some average number $\bar N_1$. Since this structure will play no role in the analysis of this article, we will suppress the linear combination and assume that $N_1$ is directly the average value.} generalised free field representation of the dual states is thus
\begin{equation}\label{eq:dualstate}
\left( L_{-1}^{n+1} \tilde L_{-1} \,|O_{\frac{1}{2},\frac{1}{2}}\rangle\right )^{N_1} \left(|0\rangle_{NS}\right)^{N_0} \,,
\end{equation}
where $|0\rangle_{NS}$ is the SL(2,$\mathbb{C}$)-invariant vacuum, and the numbers of single particle constituents are constrained to sum up to $N$: $N_0+N_1=N$.  Based on the experience with the BPS solutions (see for example \cite{Giusto:2015dfa}), we expect $\frac{N_1}{N}\sim \alpha^2$ at leading order in $\alpha$;  the precise relation will be derived below. 

We should first clarify what kinds of quantitative checks one can make to support the equivalence between the state \eqref{eq:dualstate} and the geometry constructed in section~\ref{sec:sugra}. The primary tool, introduced in \cite{Kanitscheider:2006zf,Kanitscheider:2007wq} and then used extensively in \cite{Giusto:2015dfa,Giusto:2019qig,Rawash:2021pik}, has been to compute the vacuum expectation values (VEVs) of CPOs in the microstates both in the free orbifold description of the CFT and in the dual geometry, where the VEVs are encoded in the asymptotic expansion: when the microstates preserves some supersymmetry, these VEVs do not depend on the moduli \cite{Baggio:2012rr} and the two computations must match. On the contrary the CPO VEVs in non-BPS states are not protected quantities and a quantitative comparison between supergravity and orbifold CFT is not possible. All one can do is to verify that the VEVs satisfy basic consistency conditions, to be specified more precisely in the course of our analysis, and to use the gravity result as a prediction for the strongly coupled regime of the CFT. The general pattern that will emerge is that the corrections to the free CFT results are due to the interaction between the $N_1$ elementary constituents $L_{-1}^{n+1} \tilde L_{-1} \,|O_{\frac{1}{2},\frac{1}{2}}\rangle$, and hence are proportional to powers of $G_N\,N_1\sim \alpha^2$, where $G_N\sim N^{-1}$ is the Newton's constant and we used $N_1\sim N\alpha^2$.

The first piece of information one needs for a more quantitative comparison between gravity and CFT is the precise relation between the gravity parameter $\alpha$ and the number of single particle constituents $N_1$. This could be inferred by comparing protected (i.e. moduli-independent) quantities, like the momentum charge $n_p$ or the $SU(2)_L\times SU(2)_R$ angular momenta $(j,\bar j)$. The holographic recipe to extract these charges from an asymptotically AdS geometry is given for example in \cite{Hansen:2006wu,Kanitscheider:2006zf}. One defines a radial coordinate $z$ such that at the AdS boundary $z\to 0$ the 3D Einstein metric $ds^2_3$ and the $SU(2)_L\times SU(2)_R$ gauge fields,
\begin{equation}
A^{\pm}\equiv A^{\varphi_1}\pm A^{\varphi_2},
\end{equation}
 have the expansion
\begin{equation}
\begin{aligned}
ds^2_3 &= \frac{dz^2 }{z^2}+\frac{1}{z^2} \left(g^{(0)}_{\mu\nu} +z^2 g^{(2)}_{\mu\nu} \right) dx^{\mu} dx^{\nu} + O(z^2) \,,\quad g^{(0)}_{\mu\nu} dx^\mu dx^\nu =-d\tau^2+d\sigma^2\,,\\
A^{\pm}&= A^{(0) \pm}_\mu dx^\mu+O(z^2)\,,
\end{aligned}
\end{equation}
where the indices $\mu,\nu$ range over the 2D coordinates $\tau,\sigma$. In our solutions of the $\alpha$-class the coordinates $z$ and $\rho$ are related as
\begin{equation}
\rho=\left(1+\frac{3}{8(n+1)^2}\alpha^2+c^{(0)}_\rho \alpha^4+O(\alpha^6)\right) z^{-1}\left[1-\frac{z^2}{4}\left(1+c^{(1)}_\rho\alpha^4+O(\alpha^6)\right) + O(z^4) \right]\,,
\label{eq:rhoAlpha}
\end{equation}
where $c^{(0)}_\rho$, $c^{(1)}_\rho$ are $n$-dependent rational numbers :
\begin{align}
  c^{(0)}_\rho ~&=~ \frac{72 \big(3 n (n+3)+4\big) \qty(-3 H_{n+1} + H_{n+\frac{5}{2}} + \log 4) + 3(1 + 2 n) \big(89 + 4 n (35 + 9 n)\big)}{128 (2 n+1) (2 n+3) (2 n+5)(n+1)^4} \,,
  \\
  c^{(1)}_\rho ~&=~ \frac{36 \big(3 n (n+3)+4\big) \qty( H_{n+1} - H_{n+\frac{5}{2}} - \log 4) - 24 n (6 n^2 + 13n - 12) + 231}{32 (2 n+1) (2 n+3) (2 n+5) (n+1)^4} \,,
\end{align}
and we have defined
\begin{equation}
  H_k ~=~ \int_0^1 \frac{1-x^k}{1-x} dx,
\end{equation}
which is equal to the $k$-th harmonic number when $k$ is an integer.

Note that to obtain the relations one needs the expansion of the metric up to $O(\alpha^4)$. In section \ref{sec:nonbpsalpha}, we computed the expansion of all the fields for a general $n$ up to order $O(\alpha^3)$ only. While we have not been able to obtain the exact general expression of the metric functions at order $O(\alpha^4)$, we could compute the first terms of their asymptotic expansions as $\rho \to \infty$. This in turn lets us compute the first terms of the asymptotic expansion of $\rho$, given above, in terms of the new coordinate $z$, as well as $g^{(0)}_{\mu\nu}$ and $g^{(2)}_{\mu\nu}$, up to order $O(\alpha^4)$.

The VEVs of the stress-energy tensor $T_{\mu\nu}$ and the R-symmetry currents $J_\mu$, $\tilde J_\nu$ are given in terms of the 
coefficients of the $z\to 0$ expansion\footnote{We are considering a particular case of a more general analysis~\cite{Henningson:1998gx,Balasubramanian:1999re,deHaro:2000vlm}: first we are taking the CFT spacetime to be flat $R\times S^1$ so we neglect the curvature contribution to the Weyl anomaly, then we also focus on the value of the currents away from the location of other operators so we neglect any contact terms. With respect to~\cite{Kanitscheider:2006zf,Kanitscheider:2007wq} we use the equations of motion to rewrite the contributions coming from the normalisable modes of the matter fields in terms of the trace of $g^{(2)}$.} by
\begin{equation}
\begin{aligned}
\langle T_{\mu\nu} \rangle &=  \frac{N}{2\pi} \left[g^{(2)}_{\mu\nu} -\frac{g^{(0)}_{\mu\nu} }{2}  g^{(2)\lambda}_{\lambda}+\frac{1}{8}\sum_{\pm} \left(A^{(0) \pm}_\mu  A^{(0) \pm}_\nu-\frac{g^{(0)}_{\mu\nu}}{2} A^{(0) \pm \lambda}  A^{(0) \pm}_\lambda \pm {\epsilon_{(\mu}}^{\lambda} A^{(0)\pm}_\lambda A^{(0)\pm}_{\nu)} \right)\right],\\
\langle J_{\mu} \rangle &=  \frac{N}{8\pi}\left(A^{(0) +}_\mu +{\epsilon_{\mu}}^{\nu} A^{(0) +}_\nu\right)  \,,\quad \langle \tilde J_{\mu} \rangle =   \frac{N}{8\pi}\left(A^{(0) -}_\mu-{\epsilon_{\mu}}^{\nu}  A^{(0) -}_\nu\right),
\end{aligned}
\end{equation}
where indices are raised and lowered with $g^{(0)}_{\mu\nu}$. 
Note that 
\begin{equation}
g^{(0) \mu\nu}\langle T_{\mu\nu} \rangle=0\,,\quad {\epsilon_{\mu}}^{\nu}  \langle J^{\pm}_\nu\rangle = \pm \langle J^{\pm}_\mu\rangle\,,
\end{equation}
as one expects for a CFT. These VEVs determine the left and right conformal dimensions
\begin{equation}\label{eq:hbarhhol}
\begin{aligned}
h - \frac{N}{4} &= \pi \left(\langle T_{\tau\tau}\rangle +\langle T_{\tau\sigma} \rangle\right)=\frac{N}{4} \left[g^{(2)}_{\tau\tau}+2g^{(2)}_{\tau\sigma}+g^{(2)}_{\sigma\sigma}+\frac{1}{4}(A^{(0)+}_\tau+A^{(0)+}_\sigma)^2\right]\,,\\
 \bar h - \frac{N}{4} & = \pi \left(\langle T_{\tau\tau}\rangle -\langle T_{\tau\sigma} \rangle\right)=\frac{N}{4} \left[g^{(2)}_{\tau\tau}-2g^{(2)}_{\tau\sigma}+g^{(2)}_{\sigma\sigma}+\frac{1}{4}(A^{(0)-}_\tau-A^{(0)-}_\sigma)^2\right],
\end{aligned}
\end{equation}
where the shift by $\frac{N}{4}=\frac{c}{24}$ is the NS-sector zero-point energy. The $SU(2)_L$, $SU(2)_R$ angular momenta $j$, $\bar j$ are given by
\begin{equation}\label{eq:holangular}
j=2\pi \,\langle J_\tau\rangle =\frac{N}{4} \,(A^{(0)+}_\tau+A^{(0)+}_\sigma) \quad,\quad \bar j =2\pi \, \langle \tilde J_\tau\rangle=\frac{N}{4}  \, (A^{(0)-}_\tau-A^{(0)-}_\sigma)\,.
\end{equation}
While $h$ and $\bar h$ depend on the moduli, the momentum charge
\begin{equation}
n_p=h-\bar h
\end{equation}
does not. We can thus use this charge as a benchmark\footnote{Note that this method does not work for $n=0$, since it would give a trivial $0=0$ relation. Also one cannot use as benchmarks the conserved angular momenta $j$, $\bar j$ for the reason explained in the next paragraph.} to establish the relation between $\alpha$ and $N_1$: this is done by equating the CFT result $n_p=N_1 n$ with the gravity answer obtained through the above holographic recipe. We obtain the equation
\begin{equation}
  g^{(2)}_{\tau\sigma} ~=~ n \frac{N_1}{N} \,,
\end{equation}
and we can compute the perturbative expansion in $\alpha$ of the left-hand side. We then invert the series to obtain an expansion of $\alpha$ in powers of $N_1/N$:
\begin{equation}\label{eq:avsN1}
\frac{\alpha^2}{(n+1)^2} = 4\frac{N_1}{N}\left[1+c_{N1} \frac{N_1}{N} + O \left(\frac{N_1}{N}\right)^2  \right]\,,
\end{equation}
where $c_{N1}$ are $n$-dependent rational numbers:
\begin{equation}
\begin{aligned}
  c_{N1} ~=~ \frac{12 \big(3 n (n+3)+4\big) \qty( 3 H_{n+1} - H_{n+\frac{5}{2}} - \log 4) + 4 n (4 n+21)+39}{2 (2 n+1) (2 n+3) (2 n+5)} \,.
\end{aligned}
\end{equation}

It might be surprising that the angular momenta depend on the asymptotic values of $A^{\pm}_\tau$, since these can be changed by a redefinition of the angular coordinates 
\begin{equation}\label{eq:largegauge}
\varphi_i\to \varphi_i+ c_{\varphi_i} \tau\,,
\end{equation}
with $c_{\varphi_i}$ some constants. The classical supergravity description is invariant under these large (i.e. non-trivial at the asymptotic boundary) gauge transformations and thus the asymptotic values of $A^{\pm}_\tau$ are completely arbitrary. In section~\ref{sec:sugra} we have arbitrarily fixed this gauge freedom by choosing the fields to depend on the ``bare'' phase $\varphi_1-\omega\tau-n \psi$, with $\omega=2m+1=3$. However, as it is explained in \cite{Hansen:2006wu}, in the quantum theory large gauge transformations are anomalous, and the change of coordinates \eqref{eq:largegauge} has the physical effect of changing the state: an example of this phenomenon is the spectral flow transformation that relates the R and NS sectors of the CFT. Hence the correct dual of the state \eqref{eq:dualstate} is obtained by applying to the solution of the previous section a large gauge transformation \eqref{eq:largegauge} where the gauge parameters $c_{\varphi_i}$ are determined by matching the holographic relations for the angular momenta \eqref{eq:holangular} with the expected CFT values $j=\bar j=\frac{N_1}{2}$. After applying the coordinate transformation \eqref{eq:largegauge} the phase appearing in the gravity solution is modified to 
\begin{equation}
\varphi_1-\omega\tau-n \psi \to \varphi_1-(\omega+\delta\omega) \tau-n \psi\,, 
\end{equation}
with $\delta\omega=-c_{\varphi_1}$. After expressing everything in terms of $\frac{N_1}{N}$ via \eqref{eq:avsN1}, we obtain a result of the form
\begin{equation}\label{eq:deltaomega}
\delta\omega=-c_{\omega}\, \frac{N_1}{N} + O\left(\frac{N_1}{N}\right)^2\,,
\end{equation}
with $c_{\omega}$ an $n$-dependent positive rational number:
\begin{equation}
  c_{\omega} ~=~ \frac{1}{8} \left(64-\frac{3}{2 n+1}-\frac{66}{2 n+3}-\frac{3}{2 n+5}\right) \,.
\end{equation}

It is natural to interpret the frequency shift $\delta\omega$ as the correction to the energy of a single particle constituents $L_{-1}^{n+1} \tilde L_{-1} \,|O_{\frac{1}{2},\frac{1}{2}}\rangle$ due to the (attractive) interaction with the other elements of the bound state: this correction should be proportional to the number of elements $N_1$ (for $N_1\gg 1$) and to the Newton's constant $G_N\sim N^{-1}$, in agreement with \eqref{eq:deltaomega}. To check the numerical coefficient $c_\omega$ would require to compute an anomalous dimension of the operator $L_{-1}^{n+1} \tilde L_{-1} \,O_{\frac{1}{2},\frac{1}{2}}$ in the strongly coupled CFT, a task that is at the moment out of reach. We will further comment on this issue in the concluding section. 

Even without an independent verification of \eqref{eq:deltaomega}, we can still make a non-trivial check of our interpretation of $\delta\omega$ and of the consistency of our supergravity construction by comparing with the energy $h+\bar h$ extracted from the metric via the holographic relations \eqref{eq:hbarhhol}. Indeed, using these relations we compute the energy at order $\mathcal{O}(\alpha^4)$ and find 
\begin{equation}\label{eq:checkenergy}
h+\bar h = N_1 \left[(3+n) +\frac{\delta\omega}{2}  + O\left(\frac{N_1}{N}\right)^2\right]\,. 
\end{equation}
On the other hand, this quantity represents the energy of the full state \eqref{eq:dualstate} and thus should be, in first approximation, $(3+n) N_1$ which is the sum of the free energies of the elementary constituents of the bound state. Then we should include the dynamical correction: in the regime $1\ll N_1\ll N$ this is given by the interaction energy between any single pair of constituents, times the number of pairs. But according to our interpretation of $\delta\omega$ the interaction energy between one pair should be identified with $\frac{\delta\omega}{N_1}$, and the number of pairs is $\approx \frac{N_1^2}{2}$ for $N_1\gg 1$. This way of calculating $h+\bar{h}$ reproduces the holographic result \eqref{eq:checkenergy}, including the $N_1/N$ correction in the square parenthesis.

We have performed the analysis above for the $\beta$-class of solutions as well. The results are qualitatively the same. Unfortunately, we were unable to find closed form expressions, the $\beta$ analogues of \eqref{eq:rhoAlpha}, \eqref{eq:avsN1} and \eqref{eq:checkenergy}, valid for arbitrary $n$, even though we have explicit values for $n$ from $1$ to $80$.

The interpretation of the frequency shift $\delta\omega$ as the interaction energy between non-BPS constituents can also explain the difficulties encountered in finding within the ansatz \eqref{eq:metric6D3D}-\eqref{eq:Xphi} a geometry dual to a more generic state than \eqref{eq:dualstate} where both types of perturbations, \eqref{eq:LLO12} and \eqref{eq:LLO11}, are present at the same time. In the Q-ball ansatz the metric oscillates with a phase that is twice the fundamental one ($\varphi_1-(\omega\tau+n\psi)$) appearing in the field $X$ \eqref{eq:Xphi}: on the CFT side this implies that the energy and momentum associated with the constituents of type \eqref{eq:LLO11} have to be twice the ones carried by the constituents of type \eqref{eq:LLO12}. In the free theory this is easily engineered by linking appropriately the values of $n$ and $m$ in \eqref{eq:LLO12} and \eqref{eq:LLO11}. We have seen however that interactions might spoil this relation since there is no reason that the quantum correction to the energy of the constituents \eqref{eq:LLO11} be twice the one for the constituents \eqref{eq:LLO12}. To keep into account this effect one would need to generalise the ansatz \eqref{eq:metric6D3D}-\eqref{eq:Xphi} by introducing more than one basic frequency. Note however that this problem does not arise for BPS states, whose energy is protected, and this explains why it is possible to find regular and normalisable solutions where both the parameters $\alpha$ and $\beta$ are independently turned on within the Q-ball ansatz in the supersymmetric case \cite{Ganchev:2021new}.

Finally we comment on the VEVs of the CPOs $O_{1/2,1/2}$ and $O_{1,1}$. The VEV of $O_{1/2,1/2}$ is encoded in the asymptotic expansion of $C_0$. It is immediate to check that at order $\alpha$ the holographic result agrees with the orbifold CFT result: this is not a surprise, since the gravity solution at $O(\alpha)$ can be generated by acting with $\tilde L_{-1}$ on the BPS superstratum. The non-BPS nature of the state is visible only in the corrections of higher order in $\alpha$, where the agreement between the gravity and the free orbifold CFT result no longer holds. A qualitatively more instructive example is that of $O_{1,1}$. In the free orbifold theory its VEV vanishes, as there is an enhanced symmetry connecting it to the BPS superstratum, however there is no reason to expect that this result continues to hold away from the orbifold locus and in particular at the supergravity point. On the gravity side the VEV of $O_{1,1}$ is captured by $\mu_0$, as explained in section~\ref{sec:trunCFT}. We saw that it is crucial for having a regular and normalizable solution to add a homogeneous term of order $\alpha^2$ to $\mu_0$ and this precisely denotes the non-vanishing VEV of $O_{1,1}$; the fact that the VEV is $O(\alpha^2)$ (at leading order) suggests that the effect is again due to the interaction between the non-BPS strands whose number is controlled by $\alpha$. 
 
\section{Summary and outlook}
\label{sec:conclusions}

In this paper we provided evidence that multi-particle states, where the number of single-particle BPS constituents is a finite fraction of $N$ in the $N\to \infty$ limit, are described by regular asymptotically AdS geometries even when the state does not preserve globally any supersymmetry. These states are non-BPS analogues of the superstrata constructed in \cite{Bena:2015bea,Bena:2016ypk} and share with them many qualitative features but they also display new dynamical properties. They are due to the interactions between the single-particle constituents, which, by themselves, are descendants of chiral primaries. In the absence of the linear structure which simplifies the BPS analysis, the main technical tool that allows the construction of the non-BPS solutions is the 3D truncation and the Q-ball ansatz introduced in this context in \cite{Mayerson:2020tcl,Houppe:2020oqp,Ganchev:2021pgs}. Even within this ansatz, an analytical approach is for now viable only in a perturbative expansion in the ratio between the number of single-particle constituents and $N$: though we can only present our results up to a finite perturbative order, we have accumulated evidence that the construction works at arbitrary order. It is crucial for the existence of regular solutions to exploit all the degrees of freedom contained in the truncation and, in particular, we have pointed out that if one turns on the $\nu$-perturbation \eqref{eq:nu} at linear order, one has to fine-tune the coefficient of the other independent ($\mu_0$) perturbation \eqref{eq:mu0} at the higher orders to ensure regularity and not to spoil the asymptotic behaviour. This is a form of ``coiffuring'' for which we provide a CFT interpretation by mapping the coefficient of the $\mu_0$-perturbation to the expectation value of a certain CPO \eqref{eq:O11mu0}, which is non-vanishing in the non-BPS state at the supergravity point. 

Despite the success of the 3D truncation and the Q-ball ansatz at describing a class of non-BPS states, more general states will require a generalisation of the ansatz. For example, non-BPS ``multi-mode'' states containing at the same time single-particle constituents of the type \eqref{eq:LLO12} and \eqref{eq:LLO11} cannot be described within the ansatz. The root of this difficulty lies in the interactions that modify the energy of the single-particle constituents away from their free field values: while in the generalised free theory the state \eqref{eq:LLO11} has twice the energy of the state \eqref{eq:LLO12}, this is no longer true in the interacting gravity theory. Since energy is related to the phase of the fields, and the ansatz \eqref{eq:metric6D3D} allows only phases that are integer multiples of one basic phase, a general non-BPS state with both types of single-particle constituents is not described by our ansatz. This problem does not arise for BPS states, where the energy is protected, and indeed a supersymmetric solution where one turns on both the $\nu$ and the $\mu_0$ perturbations has been found in \cite{Ganchev:2021new}. We emphasise that the obstacle one encounters in the construction of the multi-mode non-BPS solutions is merely a technical problem related with a too restrictive choice of ansatz, and does not indicate that such states do not exist in the interacting theory.

Another limitation of our current analysis lies in the perturbative nature of our gravity construction. The solutions we can analytically construct  are small perturbations around the AdS$_3\times S^3$ vacuum and as such are far from the black hole regime. To describe typical non-BPS states of the D1-D5-P black hole it would be useful to study the evolution of a deep supersymmetric microstate when it is perturbed away from extremality and we hope that this could be done via a multi-mode generalisation of the Q-ball ansatz. Linked to this question is the problem of extending the non-BPS solutions from the asymptotically AdS region to the asymptotically flat region. This is equivalent to coupling the branes degrees of freedom described by the CFT to the closed strings and through this coupling the non-BPS states are expected to radiate away their excess energy, as it was observed, at the linearised level, in the analysis of \cite{Bombini:2017got}. Thus the knowledge of the asymptotically flat solution should reveal the full dynamics of the non-BPS states. 

Finally it would be worthwhile to explore further the CFT interpretation of the non-BPS solution we have obtained. Given the absence of non-renormalisation theorems, a precise holographic analysis like the one performed in \cite{Kanitscheider:2006zf,Kanitscheider:2007wq,Giusto:2015dfa,Giusto:2019qig,Rawash:2021pik} for the supersymmetric microstates is likely out of reach. Nevertheless, the holographic dictionary reviewed in section~\ref{sec:holo} allows us to extract from the geometry predictions for the VEVs of CPOs in the non-BPS state as well as for the average dimensions of the multi-particle operators made with the constituents in \eqref{eq:LLO12}, \eqref{eq:LLO11}. For instance the $\delta\omega$ given in \eqref{eq:deltaomega} should be linked with the anomalous dimension of a two-particle operators of the form $(L_{-1}^{n+1} \tilde L_{-1} O_{\frac{1}{2},\frac{1}{2}})^2$. The information about such anomalous dimension is in principle contained in the four-point correlator involving the CPO $O_{\frac{1}{2},\frac{1}{2}}$, which has been computed in \cite{Giusto:2018ovt}. In the AdS$_3$ context anomalous dimensions of two-particle operators have already been derived in \cite{Aprile:2021mvq,Ceplak:2021wzz}. However finding a precise relation between CFT anomalous dimensions and our non-BPS supergravity solutions is technically challenging. First the operator $(L_{-1}^{n+1} \tilde L_{-1} O_{\frac{1}{2},\frac{1}{2}})^2$ is a mixture of primaries and descendants which will develop different anomalous dimensions as the interaction between the two constituents is switched on. Then each primary involved in the double particle state above will actually be a linear combination of constituents with different flavours. In the generalised free theory the various representations of the flavour group have the same dimension, but again develop different anomalous dimension in the interacting case. In order to resolve this flavour mixing one would need to know four-point correlators involving CPOs other than $O_{\frac{1}{2},\frac{1}{2}}$, which are currently not available explicitly. However, we think that, as new results become available, it is likely that the dynamical information encoded by non-BPS solutions such as the ones discussed here will provide new data also for the strongly coupled D1-D5 CFT, as it happened for the BPS case~\cite{Giusto:2018ovt}.

\section*{Acknowledgements}

We would like to thank Nicholas Warner for stimulating discussions throughout the creation of this work, Daniel Mayerson for helpful comments on the 10D uplift and Kostas Skenderis for a clarifying email exchange about the definition of the holographic stress tensor.
This work was partially supported by the Science and Technology Facilities Council (STFC) Consolidated Grant ST/T000686/1 “Amplitudes, strings \& duality”. The work of BG and AH is supported in part by the ERC Grant 787320 - QBH Structure. The work of SG was supported in part by the Research Funds, F.R.A. 2019, of the Department of Physics of the University of Genoa and by the MIUR-PRIN contract 2017CC72MK003.

\appendix

\section{The 10D uplift} 
\label{app:uplift10}

In this Appendix we work out the exact ten-dimensional uplift of all the six-dimensional fields. This work is mostly done in \cite{Mayerson:2020tcl} for all the fields but the three-forms. The latter can be found in the Appendix B of \cite{deLange:2015str}, albeit with different notations and conventions.

The bosonic field content of type-IIB supergravity comprises a 10D metric, a dilaton $\Phi$, the NS-NS $B_{(2)}$ field, and the Ramond-Ramond forms $C_{(0)}$, $C_{(2)}$, and $C_{(4)}$. After reducing on a torus, the term proportional to the volume of the torus in $C_{(4)}$ gives rise to a scalar named $\chi_2$, such that
\begin{equation}
  C_{(4)} ~=~ \chi_2 \,\text{vol}_{T^4} + \dots\,,
\end{equation}
and the metric decomposes into:
\begin{equation}
    ds_{10}^2 ~=~ e^{\Phi/2} \qty(e^{\phi_2/2} ds_6^2 + e^{-\phi_2/2} ds_{T^4}^2) \,,
\end{equation}
where $ds_{T^4}$ is the flat metric on the torus.

On the other hand, the six-dimensional theory of interest in this paper and in \cite{Mayerson:2020tcl} contains a metric, two scalars $X$ and $\Delta$, as well as three 3-forms $G^I$ respecting the self-duality rules
\begin{equation}
    \eta_{IJ} \star_6 G^J = - \cM_{IJ} G^J \,, \quad \cM = \mqty( 
    \frac{Q_1(X^2 + 2 \Delta)^2}{4Q_5 \Delta} & \frac{X^2}{2\Delta} &  \sqrt{\frac{Q_1}{2Q_5}} \frac{X(X^2 + 2 \Delta)}{2\Delta}
    \\
    \frac{X^2}{2\Delta} & \frac{Q_5}{Q_1 \Delta} &  \sqrt{\frac{Q_5}{2 Q_1}} \frac{X}{\Delta}
    \\
     \sqrt{\frac{Q_1}{2Q_5}} \frac{X(X^2 + 2 \Delta)}{2\Delta} &  \sqrt{\frac{Q_5}{2 Q_1}} \frac{X}{\Delta} & \frac{X^2 + \Delta}{2\Delta}
    ) \,,
    \label{eq:dualityso12}
\end{equation}
where $I=1,2,4$, and the indices are raised and lowered using
\begin{equation}
    \eta = \mqty(0 & 1 & 0 \\ 1 & 0 & 0 \\ 0 & 0 & -1/2) \,.
\end{equation}

These fields can be uplifted to the ten-dimensional theory. We start with the uplift of the scalars to 10D. They are given by:
\begin{align}
  e^{2\Phi} = \frac{Q_1}{4 Q_5} \frac{(2 \Delta+X^2)^2}{\Delta}\,&,\quad C_{(0)} = \sqrt{\frac{2 Q_5}{Q_1}}\, \frac{X}{2 \Delta+X^2}\,,
  \\
   e^{2\phi_2} ~=~ \frac{Q_5}{Q_1}\, \frac1\Delta \,&, \quad \chi_2 ~=~ \sqrt{\frac{Q_1}{2Q_5}} X \,.
\end{align}
We are left with the uplift of the 3-forms:
\begin{equation}
\begin{aligned}
    G^1 &~=~ \frac{1}2 dC_{(2)}\, , \\
    G^2 &~=~ -\frac{1}{2} e^{\Phi-\phi_2}\star_6 F_{(3)} +\frac{1}{2}\chi_2\,dB_{(2)} \equiv \frac{1}{2} d\tilde C_{(2)}\,,\\
    G^4 &~=~  - dB_{(2)} \,,
\end{aligned}
\end{equation}
where $\tilde C_{(2)}$ is the 2-form dual to $C_{(2)}$.
Because of the condition \eqref{eq:dualityso12}, the 10D forms satisfy the following duality relation:
\begin{equation}
  d B_{(2)} - \star_6 d B_{(2)} ~=~ \frac{C_{(0)}}{C_{(0)}^{\hspace{.1em}2} + e^{-2 \Phi}} \qty(d C_{(2)} - \star_6 d C_{(2)}) \,.
\end{equation}

\section{The linearized equations of motion}
\label{app:linearized_oem}

In this Appendix, we recall the expressions of the linearized equations of motion around the AdS vacuum (\ref{eq:ads_vacuum}), that were given in \cite{Ganchev:2021pgs} :

\begin{equation}
    \frac1\xi \, \partial_\xi \qty(\xi\, \qty(1-\xi^2)\, \delta\nu')  - 4\left(\frac{1}{\xi^2}- (m+1)(m+2) \, \right)\delta \nu ~=~ 0\,,
\end{equation}

\begin{spreadlines}{1em}
\begin{align}
    \frac1\xi \,\partial_\xi \qty(\xi\, \delta\mu_0') ~+~  \frac{ 4\, ( 2m + 3)^2 }{(1-\xi^2) }  \, \delta\mu_0  ~- \frac{16}{\xi^2 \,(1-\xi^2) } \, \delta\mu_0    ~=~ 0, \\
    \frac1\xi \,\partial_\xi \qty(\xi\, \delta\mu_+') \,-\, \frac{8}{(1-\xi^2)^2} \delta\mu_+ ~=~ 0 \,, \qquad 
    \frac1\xi \,\partial_\xi \qty(\xi\, \delta\mu_-') ~=~ 0,
\end{align}
\end{spreadlines}

\begin{spreadlines}{1em}
\begin{align}
& \partial_\xi \qty( \xi \delta\Phi'_+)  ~=~ 0  \,, \qquad 
 \partial_\xi \qty(\frac{1-\xi^2}{2\,\xi} \delta\Psi'_+ \,-\, \delta\Psi_+ \,-\, \delta\Phi_+) ~=~ 0 \\
& \partial_\xi \qty(\frac{1-\xi^2}{2\, \xi} \delta\Psi'_- \,+\, \delta\Psi_- \,+\, \delta\Phi_-) ~=~ 0 \,, \quad
    \partial_\xi \qty( \xi \delta\Phi'_- ) \,+\, 4 \, \delta\Psi'_- - 4\, \frac{\xi^2}{1-\xi^2}\, \delta\Phi'_-  ~=~ 0,
\end{align}
\end{spreadlines}

\begin{spreadlines}{1em}
\begin{align}
    \partial_\xi \qty(\xi \,\qty(1-\xi^2)\, \delta\Omega_1' \,-\, 2 \, \delta\Omega_1) ~&=~ 0 \,,  \\
    \partial_\xi \qty( \xi \delta\Omega_0') \,-\, 8 \frac{\xi}{\qty(1-\xi^2)^2}\, \delta\Omega_0  ~&=~ -3 \, \frac{1+\xi^2}{1-\xi^2} \, \delta\Omega_1'  \,,  \\
    \partial_\xi \qty(\frac{\delta k}{1-\xi^2}) + \frac{1+\xi^2}{1-\xi^2} \delta\Omega_1' ~&=~ \frac{2\xi}{(1-\xi^2)^2}(2 \delta \Omega_0 - \delta \Omega_1) \,.
\end{align}
\end{spreadlines}

\section{3D fields and CFT operators}
\label{app:3d-CFT}

In this Appendix we sketch a derivation of the map~\eqref{eq:O11mu0} between the $\mu_0$ field in the 3D truncation and the dual CFT operator. This can be done by using the holographic checks~\cite{Giusto:2019qig,Rawash:2021pik}. These references provided an explicit relation between the asymptotic behaviour of BPS solutions and the expectation values of dimension two operators in dual heavy states. For our purposes it is sufficient to focus on a simple 2-charge geometry which corresponds to the $(1,0,0)$ solution of Section~\ref{sec:bpsstr}. This 2-charge solution was first discussed in~\cite{Kanitscheider:2007wq} and corresponds to a profile $F_1(v') +i F_2(v')= a e^{2\pi i v'/L}$, $F_5 = -b \sin(2\pi v'/L)$. The idea is to look for a CPO of dimension two that does not have a VEV in such geometry and is a linear combination of the operators dual to a perturbation of the dilaton and the anti-delf-dual part of the RR 3-form ($s_2$), and of the 6D metric ($\tilde\sigma_2$). Then we can identify such linear combination with (the normalizable mode of) $\mu_0$ since we know that it is trivial in this geometry, see~\eqref{eq:mu10n}.

Here we follow the notation of~\cite{Rawash:2021pik} and from now we refer to the equations of that reference by using square brackets. By using the asymptotic behaviour of the $(1,0,0)$ solution, we can read the result for $s_2$ from Eq.~[D.4] of~\cite{Rawash:2021pik}
\begin{equation}
  \label{eq:fios2}
  s^{(\pm 1,\pm 1)}_2 = \sqrt{\frac{3}{2}} (f^{(\pm 1,\pm 1)}_{12} - f^{(\pm 1,\pm 1)}_{52}) = \sqrt{\frac{3}{2}} \left( \frac{a^2 b^2}{2 a_0^4} \frac{1}{2 \sqrt{3}}\right)\;.
\end{equation}
where the $f$'s are defined in Eq.~[D.2] and are related to the supergravity solution by~[C.1].

The derivation of $\tilde{\sigma}_2$ is a bit more laborious as its expression in~[D.4] requires, besides the combination $(f_{12} + f_{52}) $, also the knowledge of the asymptotic behaviour of the gauge fields. This provides a cross-check as the two contributions above separately have a non-trivial component along the spherical harmonics of spin $j=\bar{j}=0$. At the orbifold point, it is clear that there are no VEVs for these values of the spin and the non-renormalization theorem for such 3-point functions~\cite{Baggio:2012rr} ensures that the same result holds in the regime where supergravity is a reliable approximation. In fact the $j=\bar{j}=0$ harmonic in~[D.4] cancel for the $(1,0,0)$ solution when summing the various contributions and the only non-trivial terms are
\begin{equation}
  \label{eq:fiosigma2}
  \tilde{\sigma}^{(\pm 1,\pm 1)}_2 = \frac{\sqrt{3}}{4 \sqrt{2}} \frac{a^2 b^2}{a_0^4}\;.
\end{equation}
From~\eqref{eq:fios2} and~\eqref{eq:fiosigma2}, we see that the combination~\eqref{eq:O11mu0} vanishes. Thus we should identify it with the field $\mu_0$ since it vanishes in the solution $(1,0,0)$~\eqref{eq:mu10n}. In the 3D truncation, the various fields correspond to particular $S^3$ harmonics (since the reduction on the sphere has already been performed) and, as mentioned after~\eqref{eq:O11mu0}, $\mu_0$ corresponds to the real part of spin $j=\bar{j}=1$ component of $\sqrt{3} s_2 - \tilde\sigma_2)$. The orthogonal linear combination is encoded by the field $\mu_1$: the fact that $\mu_0$ and $\mu_1$ are orthogonal states on the CFT side is reflected in the bulk language by the fact that they are decoupled in the quadratic supergravity Lagrangian. Notice that $\mu_1$ is non-trivial even in the simple BPS solution we are considering and this corresponds to the fact that the corresponding CFT operator has a non-trivial VEV given by the sum of~\eqref{eq:fios2} and~\eqref{eq:fiosigma2}.

\bibliographystyle{utphys}
\bibliography{microstates2.bib}
\end{document}